\documentclass[journal]{new-aiaa}
\usepackage[utf8]{inputenc}
\usepackage[version=4]{mhchem}
\usepackage{siunitx}
\usepackage{textcomp}
\usepackage{subfigure}
\usepackage{color}
\usepackage{booktabs}
\usepackage{trimclip}
\usepackage{longtable}
\usepackage{needspace}
\usepackage{array}

\DeclareMathOperator{\sign}{sgn}
\DeclareMathOperator{\derivd}{d}
\def\AR{\text{\itshape\clipbox{0pt 0pt .32em 0pt}\AE\kern-.30emR}}

\begin{document}

\title{Applying inviscid linear unsteady lifting-line theory to viscous large-amplitude problems}

\author{Hugh J. A. Bird,\footnote{Graduate Researcher, h.bird.1@research.gla.ac.uk, Student Member AIAA} Kiran Ramesh\footnote{Lecturer, Aerospace Science Division, School of Engineering, Kiran.Ramesh@glasgow.ac.uk, Member AIAA}}
\affil{Aerospace Sciences Division, School of Engineering, University of Glasgow, Glasgow, United Kingdom, G12 8QQ}
\author{Sh\={u}ji \={O}tomo \footnote{Ph.D Student, School of Engineering, Institute for Energy Systems, s.otomo@ed.ac.uk}}
\author{Ignazio Maria Viola\footnote{Reader, School of Engineering, Institute for Energy Systems, i.m.viola@ed.ac.uk}} 
\affil{School of Engineering, Institute for Energy Systems, University of Edinburgh, Edinburgh, United Kingdom, EH9 3BF}

\date{Received: date / Accepted: date}

\maketitle

\begin{abstract}

Unsteady Lifting-Line Theory (ULLT) is a low order method capable of modeling  
interacting unsteady and finite wing effects at low computational cost.
Most formulations of the method assume inviscid flow and small amplitudes.
Whilst these assumptions might be suitable for 
small-amplitude aeroelastic problems at high Reynolds numbers, modern
engineering applications increasingly involve lower Reynolds numbers,
large amplitude kinematics and vortex structures that lead to
aerodynamic non-linearities.
 
 This paper establishes that ULLT still provides
 a good solution for low Reynolds number, large-amplitude kinematics
problems, by comparing ULLT results against those of
experimentally validated computational fluid dynamics simulations at Re=\num{10000}.
Three-dimensional (3D) effects stabilize
 Leading Edge Vortex (LEV) structures, resulting in a good prediction
 of whole wing force coefficients by ULLT. Whilst the 
inviscid spanwise force distributions are accurate for small-amplitude kinematics, the 
ULLT cannot model 3D vortical structures, and thus it cannot
 correctly predict the force distribution due the LEV. 
It can however predict the shedding of LEVs
 to a limited extent via the leading edge suction parameter
 criterion. This can then be used as an indicator of the usefulness
 of the force distribution results.

\end{abstract}

\setcounter{page}{1}

\section*{Nomenclature}

{\renewcommand\arraystretch{1.0}
\noindent\begin{longtable*}{@{}l @{\quad=\quad} l@{}}
$\AR$ & aspect ratio\\
$c$ & chord\\
$C_l$ & two-dimensional lift coefficient\\
$C_L$ & three-dimensional lift coefficient\\
$C_m$ & two-dimensional moment coefficient\\
$C_M$ & three-dimensional moment coefficient\\
$F$ & three-dimensional correction strength\\
$h$ & plunge displacement\\
$h_0^*$ & non-dimensional plunge amplitude\\
$k$ & chord reduced frequency\\
$K$ & three-dimensional interaction kernel\\
$\mathcal{L}$ & leading edge suction parameter\\
$s$ & semispan\\
$t$ & time\\
$T$ & oscillation period\\
$U$ & free stream velocity\\
$x,y,z$ & body centered coordinate system\\
$x_m^*$ & non-dimensional reference location for pitching moment\\
$\Gamma$ & bound circulation\\
$\zeta$ & spanwise coordinate\\
$\nu$ & span reduced frequency\\
$\phi$ & velocity potential\\
$\omega$ & angular frequency\\
\end{longtable*}}

\section{Introduction}
\label{sec:intro}

\lettrine{T}{raditionally}, unsteady aerodynamics focused on inviscid, small amplitude problems. 
This was to reduce the complexity of the problem sufficiently for analytical solutions to
be obtained using low-order models. 
These low-order models discard non-essential elements of 
the flow physics to make the problem easier to solve. 
Common models are based on potential flow and assume that the fluid is inviscid and
incompressible. Geometric assumptions are often
made in order to linearize the problem.
To formulate a low order method, the problem's physics must first be understood.

Research on the unsteady aerodynamics of finite wings
was initially driven by the problem of dynamic stall in 
helicopters~\cite{McCroskey1981}. Computational Fluid Dynamics (CFD) and experimental results are reviewed
by Carr~\cite{carr1988progress}, Carr et al.~\cite{carr2} and Ekaterinaris and Platzer~\cite{ekaterinaris1998computational}. More
recent investigations include those of Angulo et al.~\cite{andreu2019influence} and
Visbal and Garmann~\cite{visbal2019dynamic,visbal2019effect}. These studies
involved oscillating wings in the high Reynolds number,
low reduced frequency regime. 
Modern applications such as micro air vehicles, high altitude long 
endurance drones and unmanned aerial vehicles have inspired research at
lower Reynolds numbers, with greater amplitudes and
higher reduced frequencies, often involving Leading
Edge Vortices (LEVs), trailing edge vortices and wing tip 
vortices.
Studies have been carried out on translating wings~\cite{mulleners2017flow,
  mancini2015unsteady,devoria2017mechanism},
pitching wings~\cite{jantzen2014vortex,hord2016leading,yilmaz2012flow,yilmaz2010scaling,visbal2017unsteady,green2011unsteady}, 
plunging wings~\cite{calderon2013volumetric,calderon2014absence,calderon2013lift,visbal2013three,yilmaz2010three,fishman2017structure},
rotating wings~\cite{ozen2012three,carr2013finite,carr2015aspect,medina2016leading,beals2015lift,venkata2013leading}, and wings subject to 
gusts~\cite{perrotta2017unsteady,corkery2018development,biler2019experimental}, and have shed light on  contributions from 
circulatory, apparent mass and vortical effects.

In two dimensions (2D), discrete vortex particle methods have proven valuable in low-order modeling. 
Linear potential flow theories such as Theodorsen~\cite{Theodorsen1935} assume small-amplitude
oscillations and a planar wake. By using discrete vortex particles to model the non-planar wake, Ramesh et al.~\cite{Ramesh2013} 
and Yan et al.~\cite{Yan2014} considered arbitrarily large kinematics. 
Criteria for the shedding of LEVs have allowed
these models to be extended to include intermittent LEV shedding. The 
Leading Edge Suction Parameter (LESP) hypothesis~\cite{Ramesh2014} states that for a given flow regime,
an airfoil can only provide so much suction at the leading edge. If this suction is exceeded,
vorticity is shed from the leading edge. This suction was initially
linked to the vorticity distribution of a thin airfoil model by Katz~\cite{Katz1981}
and Ramesh~\cite{Ramesh2014}.

Less progress in low-order modeling has been made in three dimensions (3D). 
Boundary element methods~\cite{moored2018unsteady}, 
unsteady vortex lattice methods~\cite{Katz2001, Murua2012a,Smyth2019,Hirato2019} 
or vortex particle methods~\cite{Willis2007} can be
used. However, these methods lack robust and simple LEV shedding
criteria, and have numerical difficulties in modeling unsteady wakes.
This shortcoming of 3D methods can be overcome with strip theory~\cite{Leishman2006}. 
Multiple 2D solutions can be applied to the varying chord
distribution of a 3D wing. But this neglects important 3D effects,
failing to account for a loss of lift due to induced downwash in low aspect ratio wings, 
for example. By correcting these multiple unsteady 2D solutions to account
for 3D effects, Unsteady Lifting-Line Theory (ULLT) is obtained.

ULLT uses the same ideas as steady lifting-line theory, originally credited
to Prandtl~\cite{Prandtl1923}. Prandtl combined 2D models with a simplified 3D correction
model to correct for finite wing effects. This idea of separating a difficult problem
into two simple problems with different length scales was formalized by 
Van Dyke~\cite{Dyke1964}. 

ULLT  includes additional corrections for variations in vorticity downstream
of the wing due to the unsteady nature of the problem.
Early work~\cite{James1975,Holten1976}
was asymptotically limited to low-frequency oscillation of the wing.
Further work~\cite{Ahmadi1985,Sclavounos1987} extended the frequency range before Guermond and Sellier~\cite{Guermond1991}
produced a method suitable for swept wings at any oscillation frequency.
Research in the time domain has been less extensive, and asymptotically limited
by the assumptions made in modeling the wake \cite{Jones1940,Devinant1998,Ramesh2017,Boutet2018,Sugar-Gabor2018,Bird2019}.
A more detailed review can be found in Bird and Ramesh~\cite{Bird2021b}.

These ULLT are all based upon potential-flow theory. They assume
that the flow is inviscid and incompressible. They also all assume that
vorticity is only shed from the trailing edge of the wing, and cannot model
LEVs. However, in practice, the theoretical limitations of a model
can belie its applicability.

For LLT, this has been demonstrated repeatedly: the asymptotic
limitations of the method mean that it is not strictly applicable to 
elliptic or rectangular wings, as explained by Van Dyke~\cite{Dyke1964}. 
In practice, it provides good predictions. 
However, doing so may require discretization to avoid problems with wingtip
downwash singularities. In ULLT, Bird and Ramesh~\cite{Bird2021b} found that
simplified wake models were often sufficient to obtain a good
solution. And in 2D, McGowan et al.~\cite{McGowan2011} found that Theodorsen's
method could provide good solutions even at low Reynolds number
and in the presence of LEV shedding when examining the problem of pitch-heave
cancellation. 
Similar results for more
complex kinematics can be found in Elfering and Granlund~\cite{Elfering2020},
and for finite wings in Bird and Ramesh~\cite{Bird2018}.

This paper aims to explore the limits of applicability of unsteady lifting-line
theory when applied to low Reynolds number flows and large-amplitude
kinematics. Low order models that assume inviscid flow have frequently been
successful for for finding lift and moment coefficients in the low
Reynolds number regime. Verifying that ULLT formulated for 
inviscid problems can be applied to such cases is a stepping
stone to the problem of accounting for LEV shedding. 

An unsteady lifting-line theory, based on Sclavounos' method~\cite{Sclavounos1987,Bird2021b}, will be applied to oscillating heave problems in 
the low Reynolds number regime ($Re=$\num{10000}) for wings of aspect ratios 6, 3 and 1. Low amplitude,
LEV-free cases will be examined along with larger amplitude cases. Through
comparison with experimentally validated, high-fidelity CFD results, the extent to which ULLT is usable
to obtain forces and force distributions on a wing will be established, and a 
new concept of applying the LESP criterion to predict LEV shedding using 
ULLT on finite wings will be explored.

A short introduction to Sclavounos' ULLT is given in  Sec.~\ref{sec:theory}, along with the 
means to obtain both force and leading-edge suction distributions in  Sec.~\ref{sec:theory_forces} 
and Sec.~\ref{sec:leading_edge_suction} respectively. This is followed by a discussion of the 
method's soon to be violated theoretical limitations in Sec.~\ref{sec:theoretical_limitations}.
Next, the applicability of the ULLT is explored through comparison to CFD in  Sec.~\ref{sec:results}. 
The test cases, the CFD and the validation of the CFD are detailed in 
Sec.~\ref{sec:cases},  Sec.~\ref{sec:cfd} and Sec.~\ref{sec:cfd_validation}.
This then allows the whole wing forces and LESP distributions
to be examined in  Sec.~\ref{sec:whole_wing_forces}, along with the form of the LEVs at the wing
center in  Sec.~\ref{sec:centre_wing_vorts}. Finally, the force 
distributions will be compared in  Sec.~\ref{sec:force_distributions}. Conclusions are made in 
 Sec.~\ref{sec:conclusions}.

\section{Theoretical approach}
\label{sec:theory}

In this section we present an unsteady lifting-line theory, based
on the work of Sclavounos~\cite{Sclavounos1987}. Here, we consider
pure plunge oscillations. For a more detailed derivation including pitching kinematics, 
see Bird and Ramesh~\cite{Bird2021b}.

This ULLT considers a wing in inviscid, incompressible flow
undergoing small amplitude oscillation. The freestream velocity of the 
flow is $U$ in the positive $x$ direction, and the plunge displacement of the wing is 

\begin{equation}
 	h(y; t) = h_0e^{i \omega t} = h^*_0(y)c(y)e^{i \omega t} ,
	\label{eqn:kinem}
\end{equation}

\noindent where $h$ is the plunge displacement, $h_0$ the plunge amplitude,
and $h_0^*$ is the plunge amplitude non-dimensionalized by the chord $c$. The spanwise
coordinate is $y$, $t$ is time, and $\omega$ is the angular frequency of oscillation. 
The frequency of oscillation can be non-dimensionalized
either with respect to chord or span as

\begin{equation}
	 k(y) = \frac{\omega c(y)}{2U}, \qquad \nu = \frac{\omega s}{U},
\end{equation}

\noindent where $k$ is referred as chord reduced frequency and
$\nu$ is the span reduced frequency based on the semispan $s$.

Lifting-line theory is based upon the separation of length scales.
For a wing of high aspect ratio, the chord scale is much smaller
than the span scale. This means that the detail of the chord scale
problem can be neglected in the span scale problem. And
if the chord scale problem changes slowly with respect to span, 
it can be modeled as a 2D problem with
a 3D correction. 

Here, the velocity potential $\phi(x,y,z;t)$ derived as a 2D solution with a 3D correction
is given as

\begin{equation}
 	\phi(x,y,z;t) \approx \phi^{2D}(x,z;t) + F(y)(i\omega z e^{i \omega t} - \phi^{2D}_{n}(x,z;t)),
 	\label{eq:phi_llt}
\end{equation}

\noindent where the first term $\phi^{2D}(x,z;t)$ represents the velocity potential solution to the 2D problem (which is identical to Theodorsen's problem). The
second term represents the 3D `correction' to the 2D
problem. Here, this is modeled by an oscillating uniform downwash $F(y) i \omega z e^{i\omega t}$,
and the reaction of the 2D section $F(y)\phi_n^{2D}(x,z;t)$ to this. The $F(y)$ term represents
the complex amplitude of this correction with respect to span, and 
$\phi_n^{2D}(x,z;t)$ the 2D velocity potential solution of a section oscillating in heave
with unit amplitude.

Since $\phi^{2D}$ is equivalent to Theodorsen's problem it allows the 2D bound circulation $\Gamma^{2D}$,
to be obtained:

\begin{equation}
	\Gamma^{2D}(y; t) = \frac{4 U h^*_0(y)c(y) e^{-ik}}{i H_0^{(2)}(k) + H_1^{(2)}(k)}e^{i \omega t} ,
	\label{eq:bound_vort_2D}
 \end{equation}
\noindent where $H_0^{(2)}(z)$ and $H_1^{(2)}(z)$ are Hankel 
functions of the second kind. 

Since the system is linear with respect to amplitude, this allows corrected distributions to be 
found as
\begin{equation}
	\Gamma(y; t) = \Gamma^{2D}(y;t) - F(y)\Gamma^{2D}_n(t) ,
	\label{eq:bound_vort_w_3D_correction}
\end{equation}
 where subscript $n$ once again indicates normalization for unit amplitude
$h=1$.

The 3D correction strength $F(y)$ is given by Sclavounos as

\begin{equation}
	F(y) = -\frac{1}{2 \pi i \omega e^{i \omega t}} \int_{-s}^s \Gamma'(\eta)K(y - \eta)\derivd \eta,
        \label{eqn:match2}
\end{equation}

\noindent where
\begin{equation}
	\label{eq: K unsteady}
	K(y) = \frac{1}{2s}\sign(y^*)\left[\frac{e^{-\nu |y^*|}}{|y^*|} - i\nu E_1(\nu|y^*|) + \nu P(\nu|y^*|) \right],
\end{equation}
where $E_1(x)$ is the exponential integral~\cite{Olver2010}, $y^*=y/s$ is the normalized span coordinate and 
\begin{equation}
 	P(y) = \int^\infty_1 e^{-yt}\left[ \frac{\sqrt{t^2 - 1} - t}{t}\right]\derivd t  + i \int^1_0 e^{-yt}\left[\frac{\sqrt{1-t^2}-1}{t}\right] \derivd t.
\end{equation}

Substituting this into Eq.~\ref{eq:bound_vort_w_3D_correction} gives a differential equation

\begin{equation}
  \label{eq: integro-diff equation}
  \Gamma (y; t) - \frac{\Gamma^{2D}_{n}(t)}{2 \pi i \omega} \int^s_{-s} \Gamma'(\eta)K(y-\eta) \derivd\eta = \Gamma^{2D}(y,t),
\end{equation}

\noindent where an approximate solution can be obtained by taking the bound vorticity distribution as a
Fourier series
\begin{equation}
	\Gamma(y; t) = 4Us \sum_{m=1}^M \Gamma_{m} \sin (m\zeta) e^{i\omega t},
\end{equation}
\noindent where $y = -s \cos\zeta$. This is solved at collocation points distributed over the span.

\subsection{Obtaining lift and moment coefficients}
\label{sec:theory_forces}

As with the bound vorticity, Theodorsen's theory~\cite{Theodorsen1935} allows the lift and moment
coefficients associated with $\phi^{2D}$ to be found as
\begin{align}
	C_{l}^{2D}(y;t) &= 2\pi h^*_0 (-2ikC(k) + k^2 ) e^{i \omega t} \label{eq:2d_C_l_h},\\
	C_{m}^{2D}(y;t) &= 2\pi h^*_0 \left[ - 2ik C(k) \left(x^*_m-\frac{1}{4}\right) + k^2\left(x^*_m-\frac{1}{2}\right)\right] e^{i \omega t},   \label{eq:2d_C_m_h} 
 \end{align}
\noindent where $C(k) = \frac{K_1(ik)}{K_1(ik) + K_0(ik)}$ is Theodorsen's 
function in terms of modified Bessel functions of the second kind~\cite{Olver2010}
and $x^*_m$ is the moment reference location in $[0, 1]$,
with $0$ being the leading edge and $1$ the trailing edge.

These can be corrected for 3D effects as
\begin{align}
	C_{l}(y;t) &= C_{l}^{2D}(y,t) - F(y)C_{ln}^{2D}(y), \\
	C_{m}(y;t) &= C_{m}^{2D}(y,t) - F(y)C_{mn}^{2D}(y) .
\end{align}

For the whole wing, this allow lift and moment coefficients to be found as

\begin{equation}
  C_L = \frac{1}{2s \overline{c}}\int_{-s}^{s} C_l(y)c(y)\derivd y, \qquad C_M = \frac{1}{2s \overline{c}^2}\int_{-s}^{s} C_m(y)c^2(y)\derivd y ,
\end{equation}

\noindent where $\overline{c}$ is the mean chord of the wing.

\subsection{Leading edge suction and criticality}
\label{sec:leading_edge_suction}

Recently, the idea of the leading edge suction parameter (LESP)~\cite{Ramesh2014} has become popular
as a means by which to include the shedding of a leading edge vortex
in discrete-vortex enhanced thin-airfoil theories. 

At an airfoil's rounded leading edge, suction is required for the flow to remain 
attached when going around it. The LESP is a measure of this suction, and is 
calculated from the singular leading-edge term $A_0$ in thin-airfoil theory.
For a given aerodynamic regime, it has been found that there is a maximum
LESP that can be supported by the rounded leading-edge shape. When this
critical LESP is exceeded, flow separation occurs and a shear layer is shed from the
leading edge, forming a leading edge vortex~\cite{Ramesh2014}.

For finite wings, Hirato~\cite{Hirato2020} found that the LESP criterion still
applied, with LEV formation on any 2D section occurring as predicted by
the section's leading edge suction. 
In their work, the LESP at wing section was numerically calculated using 
the unsteady vortex lattice method. In this research, we obtain an analytical
expression for the 3D LESP from the ULLT.

The inner solution of the current method is based upon Theodorsen's method.
Following Ramesh~\cite{Ramesh2020}, the leading edge suction parameter
$\mathcal{L}(y)$ can be extracted from Theodorsen's problem:
\begin{equation}
	\mathcal{L}^{2D}(y;t) = -2ikh_0^*C(k)e^{i\omega t}.
	\label{eq:lesp_2d_theodorsen}
 \end{equation}
This can then be corrected for 3D effects in the same manner as $\Gamma$, $C_l$ 
and $C_m$.
\begin{equation}
	\mathcal{L}(y;t) = \mathcal{L}^{2D}(y,t) - F(y)\mathcal{L}^{2D}_{n}(y, t) .
	\label{eq:lesp_with_respect_to_span}
\end{equation}

When $\mathcal{L}$ exceeds a certain critical value $\mathcal{L}_{crit}$ at any spanwise location, it is expected that vorticity will 
be shed from the leading edge at that section. This value $\mathcal{L}_{crit}$ must be obtained empirically.
When the critical LESP value is exceeded, we expect that the
attached flow assumption made in the derivation of the ULLT will have been violated.  
 
\subsection{Theoretical limitations}
\label{sec:theoretical_limitations}

Sclavounos' unsteady lifting-line theory is subject to theoretical
limitations both as a result of the potential flow model used, and
the asymptotic assumptions of lifting-line theory.
The derivation of both Theodorsen and Sclavounos' theories assume
inviscid, incompressible flow, a planar wake and small amplitude oscillations. It is assumed that
vorticity is shed from only the trailing edge of the wing.

Both lifting-line theory and the thin-airfoil model upon which
Theodorsen's theory is based, are asymptotic methods. The airfoil section
is assumed to be thin, and the wing is assumed to be of high
aspect ratio, with the inner (2D) solutions required to change
slowly with respect to span, in order to preserve the two-dimensionality of
the inner solution problems.
Sclavounos also assumes that the downwash induced by the outer domain
on the inner domain is uniform. This limits the asymptotic validity of the method
to high frequencies. 

In practice, users of lifting-line theory have routinely ignored
some of these theoretical limitations. Square and elliptical tipped wings
have an inner solution that changes too rapidly with respect to span near
the tip to be valid. Whilst some error is introduced near the tips,
an integro-differential approach to solving the lifting-line problem
gives a usable result in practice.

\section{Results}
\label{sec:results}

In this section, unsteady lifting-line theory is compared to results 
obtained from Reynolds-Averaged Navier-Stokes (RANS) CFD for the case of a rectangular heaving
wing. Details of the cases studied are given in Sec.~\ref{sec:cases}.
Then, the CFD method is described in Sec.~\ref{sec:cfd}. 
The CFD is validated against experiment in Sec.~\ref{sec:cfd_validation}.
Results for whole wing forces, wing center vorticity slices and 
lift distributions from both CFD and ULLT are compared in Sec.~\ref{sec:whole_wing_forces},
 Sec.~\ref{sec:centre_wing_vorts} and Sec.~\ref{sec:force_distributions} respectively.
 
 \subsection{Comparison cases}
 \label{sec:cases}
 
For all cases, a rectangular wing with a NACA0008 airfoil section and 
squared off wing tips is studied. The wing is oscillating in heave, according
to the kinematics given by Eq.~\ref{eqn:kinem}.
The Reynolds number is 10,000 and the chord reduced frequency is $k=0.4$.
The chord reduced frequency is chosen in the intermediate range since it is expected to produce
the most challenging LEV-dominated flows for the ULLT to predict. At this frequency,
3D effects remain strong and the forces are not dominated by added mass effects~\cite{Bird2021b}.
At high frequencies, added mass effects dominate the forces reducing the 3D effects~\cite{Bird2021b},
and at low frequencies strong and coherent LEVs are not expected to form.


The wing is oscillating in plunge with one of three amplitudes: $h_0^*=0.05$,
a small amplitude where leading edge vortex shedding is not expected;
$h_0^*=0.5$, an amplitude where LEV shedding is expected, and
$h_0^*=1$, a larger amplitude still. Wings of aspect ratio 6, 3 and 1
were examined along with the 2D case (\AR=$\infty$) for comparison.
These parameters are summarized in Table~\ref{tab:case_parameters}.

\begin{table}
	\centering
	\caption{Case parameters}
	\begin{tabular}{l l}
	\toprule
	Parameter & Values \\
	\midrule
	$k$ & 0.4\\
	\AR & 1, 3, 6, $\infty$\\
	$h_0^*$  & 0.05, 0.5, 1\\
	\bottomrule
	\end{tabular}
	\label{tab:case_parameters}
\end{table}

\subsection{Numerical methods}
\label{sec:cfd}

High-fidelity 3D computations of unsteady fluid dynamics are performed
at Reynolds number of \num{10000} using the open-source CFD toolbox
OpenFOAM. A body-fitted computational mesh is moved in accordance with
prescribed rate laws, and the time-dependent incompressible
Navier-Stokes are solved using a finite-volume method. A second-order
backward implicit scheme is adopted to discretize the transient terms,
while second-order, limited Gaussian integration schemes are used for
the gradient, divergence and Laplacian terms. The pressure implicit
with splitting of operators (PISO) algorithm is employed to achieve
pressure-velocity coupling. The Spalart-Allmaras (SA) turbulence
model~\cite{Spalart1992} is used for turbulence closure. The SA
model is chosen for this problem because of extensive previous
experience in applying it successfully for unsteady, separated and
vortex-dominated flows at $Re=$\num{10000} such as those considered in this
research~\cite{McGowan2011, Ramesh2014}. 
The trip terms in the original SA model are
turned off, and for the low Reynolds number cases considered in this
research, the effects of the turbulence model are confined to the shed
vortical structures and wake. 

The chord length $c= 0.1$ m. The O-mesh has
116 cells chordwise , with increased resolution near the leading and
trailing edges. There are 211, 105 and 35 cells in the spanwise
direction for aspect ratios 6, 3 and 1 respectively. The spanwise
domain extends $4$ chord lengths beyond the wingtip with an average
spacing of $21$ cells per chord length in this region. In the wall-normal direction,
cell spacing begins at $1.5\times10^{-5}$ m next to the wall ($y^+ < 1$) and extends a distance 
of $11.5$ chord lengths away from the wing with an average density of 16.3 cells per
chord length.
The simulations were
carried out at a free stream velocity $U=0.1$ m$/$s and kinematic
viscosity $10^{-6}$ m$^2/$s.

\subsection{Validation of CFD against experiment}
\label{sec:cfd_validation}
\begin{figure}
    \centering
    \subfigure[Detailed view]{
	\label{fig:expr_setup_mech}
    	\includegraphics[width=0.35\textwidth]{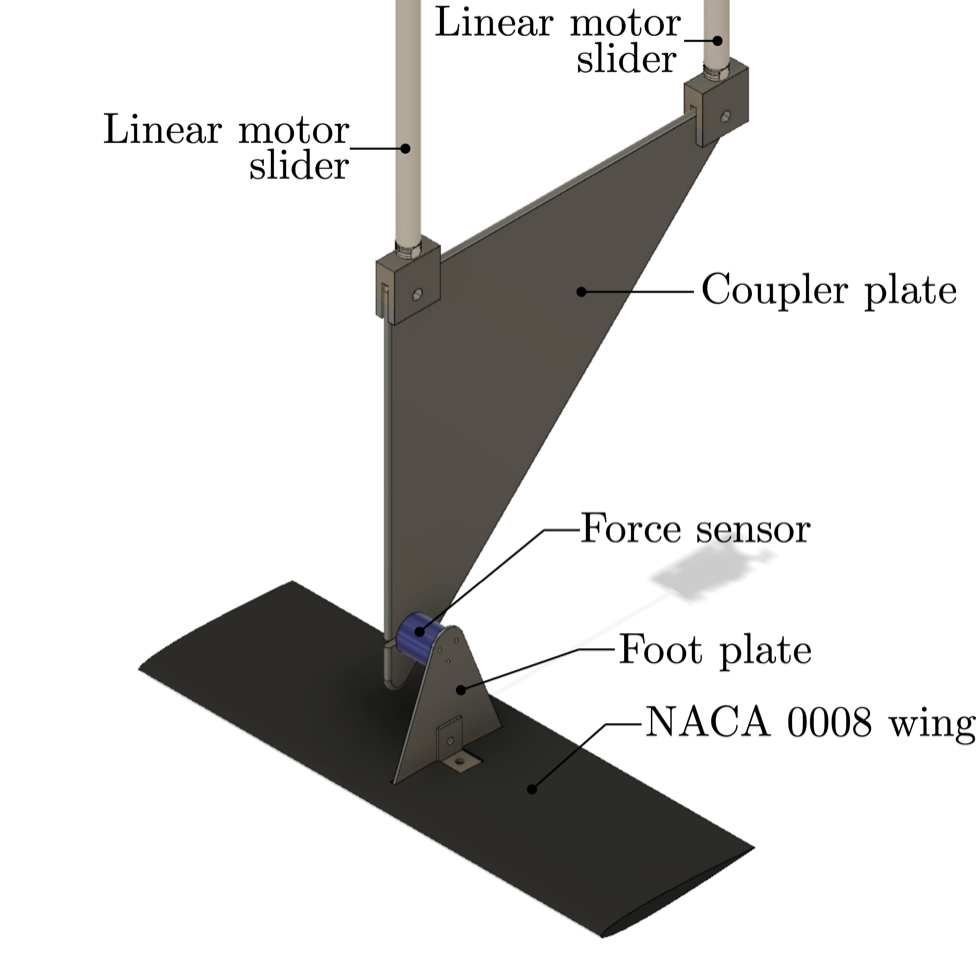}}
    \subfigure[Side view]{
	\label{fig:expr_setup_flume}
    	\includegraphics[width=0.35\textwidth]{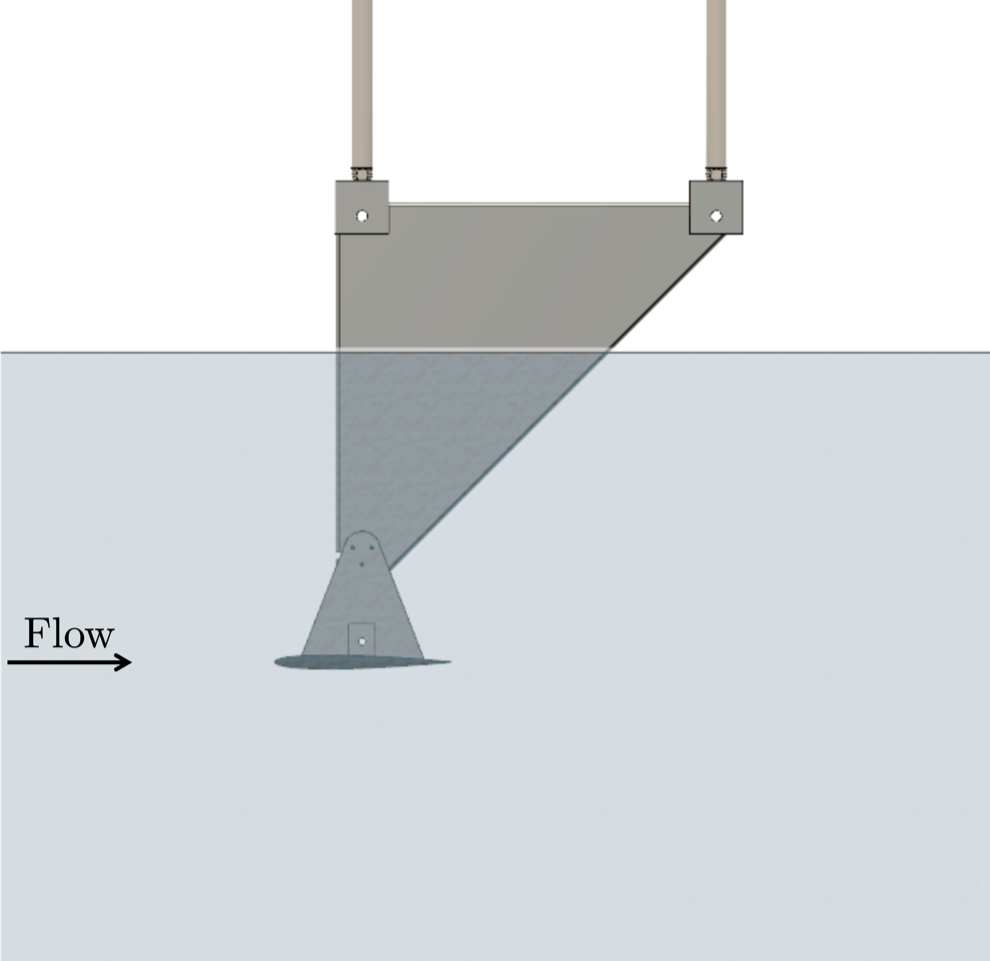}}

    \caption{Test section and experimental setup}
    \label{fig:rig}
\end{figure}

The CFD simulations were validated against an experiment performed in the water
 flume at the University of Edinburgh.
A wing with a NACA0008 airfoil section, a chord length $c$ = \SI{0.1}{\meter} and
an aspect ratio of \AR = 3 was 3D printed.
The freestream velocity was fixed at 0.1 m$/$s,
resulting in the chord based Reynolds number of \num{10000}. 
The plunging foil rig consists of two linear motors 
(LinMot, PS01-23x80F-HP-R20) connected with each other via a
 linkage system and a coupler plate. This is shown in Fig.~\ref{fig:rig}.

Direct force measurements are conducted.
 A six-axis force/torque sensor (ATI Inc., Nano-17 IP68) is mounted
 between the coupler plate and the foot plate. The sensor
is capable of measuring forces in the plane of the wing cross section up to 
$\pm$\SI{25}{\newton}, and $\pm$\SI{35}{\newton} in the orthogonal direction, and moments
 up to $\pm$\SI{250}{\newton\meter} around the three axes with a resolution of 1/160~\si{\newton} 
for the forces and 1/32~\si{\newton\meter} for the moments. LabVIEW is used to trigger 
the prescribed motor kinematics through a digital output device and also to start 
recording forces for a synchronized measurement through a DAQ board. 
A sampling frequency of 
\SI{10}{\kilo\hertz} is used to record the forces, which are then filtered in three steps. 
Firstly, a fourth-order Butterworth low-pass filter with a cutting frequency of \SI{75}{\hertz} was applied.
Then the data was smoothed with a 200 points moving average.
The last step is a sixth-order Chebyshev II low-pass filter with \SI{-20}{\decibel}
attenuation in the stopband. This three-step filtering method can preserve 
load spikes. Phase-averaging is applied for 20 periods.
 
Particle image velocimetry (PIV) is used to perform flowfield analysis .
A double pulsed Nd:YAG laser (New Wave Research, Solo PIV, \SI{532}{\nano\meter}, 
\SI{200}{\milli\joule}) is used to illuminated the plane at 1/4 of the span of the wing, 
with silver coated hollow glass spheres (Potters Industries, \SI{10}{\micro\meter})
 used as seeding. 
Images are then obtained by a CCD camera (IMPERX, B2020 equipped
 with Nikon 50~mm lens) with a resolution of 2056~pix $\times$ 2060~pix. 
Velocity vectors are computed using adaptive multi-pass cross-correlation, 
with a first interrogation window of 64~pix $\times$ 64~pix, 
and a final interrogation window of 32~pix $\times$ 32~pix, and an
 overlap of 50\% (DaVis, LaVision Inc.). Phase averaging is undertaken over 30 periods of 
Gaussian filtered PIV data.

\begin{figure}
\centering
	\label{fig:CFD_validation_cl}
    	\includegraphics[width=0.4\textwidth]{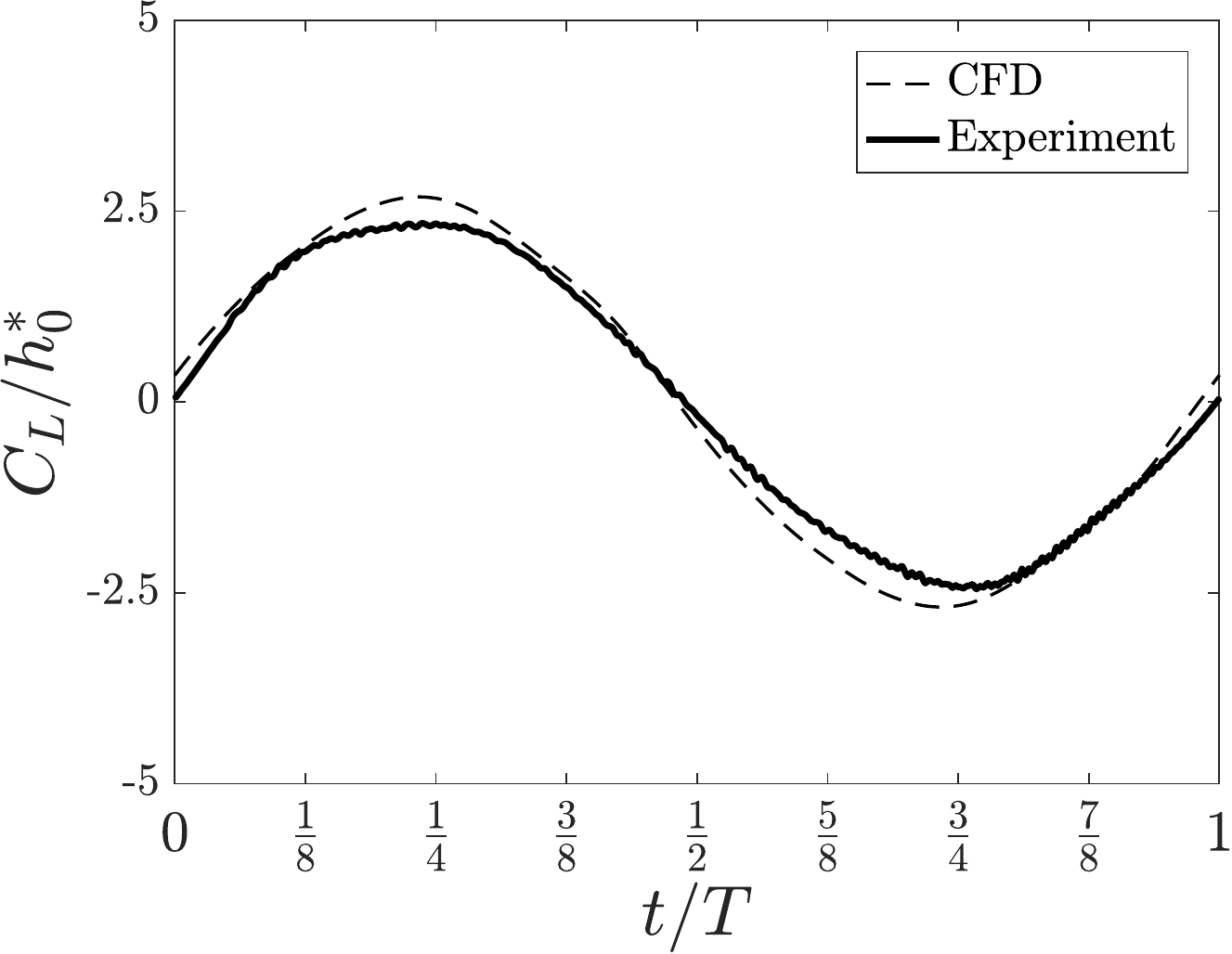}

	\caption{Comparison of the lift coefficients obtained from CFD 
	and experiment for a plunging rectangular aspect-ratio-three wing for 
	$h_0^*=0.5$ and $k=0.4$.}
	\label{fig:CFD_cl_validation}	
\end{figure}

\begin{figure}
\centering
\begin{tabular}{ >{\centering\arraybackslash}m{0.065\textwidth} >{\centering\arraybackslash}m{0.2\textwidth} >{\centering\arraybackslash}m{0.2\textwidth} >{\centering\arraybackslash}m{0.2\textwidth}  >{\centering\arraybackslash}m{0.2\textwidth}}
	\toprule
	& $t/T = 1/8$ & $t/T = 3/16 $ & $t/T =3/8$ & $t/T =1/2$ \\
	\midrule
	CFD
	&\includegraphics[width=0.21\textwidth, trim={0 0 0 8cm},clip]{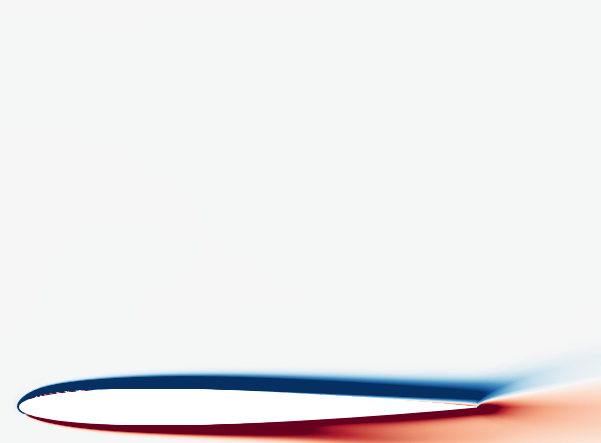}
	&\includegraphics[width=0.21\textwidth, trim={0 0 0 8cm},clip]{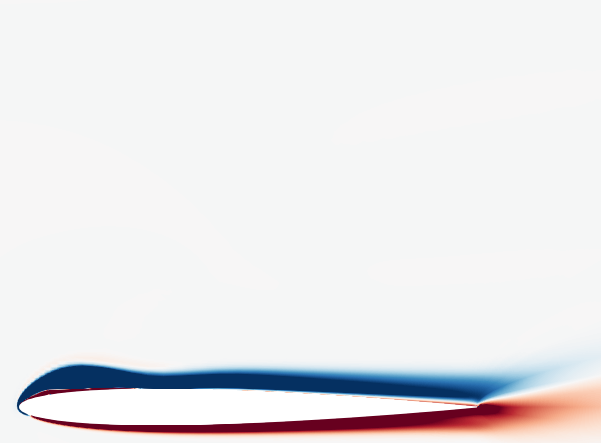}
	&\includegraphics[width=0.21\textwidth, trim={0 0 0 8cm},clip]{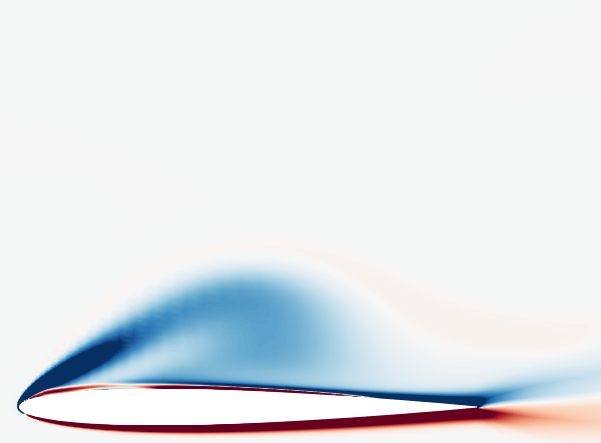}
	&\includegraphics[width=0.21\textwidth, trim={0 0 0 8cm},clip]{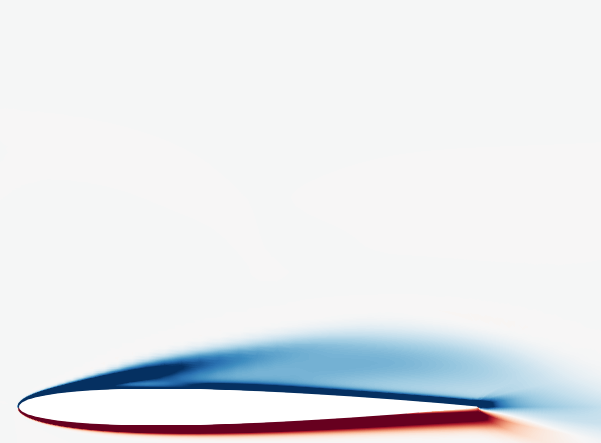}\\
	
	PIV
	&\includegraphics[width=0.21\textwidth, trim={0.8cm 3.5cm 7.5cm 0cm},clip]{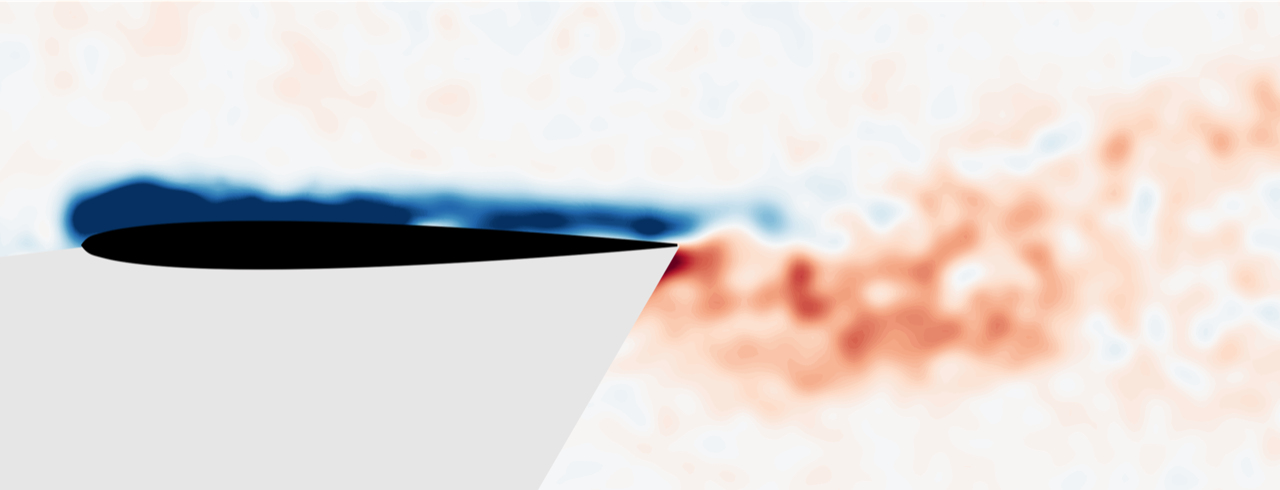}
	&\includegraphics[width=0.21\textwidth, trim={0.8cm 3.5cm 7.5cm 0cm},clip]{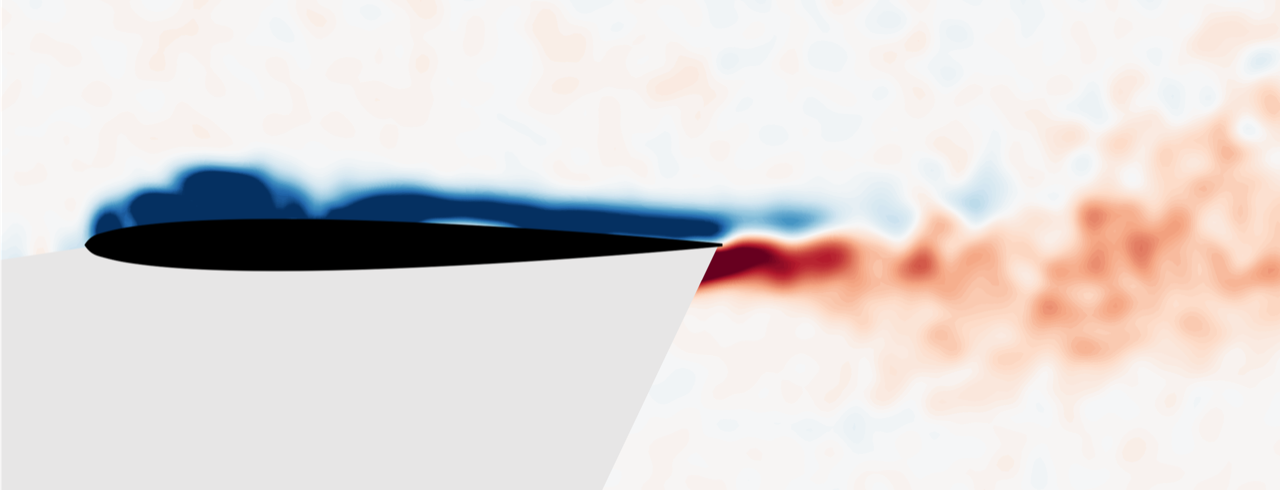}
	&\includegraphics[width=0.21\textwidth, trim={0.8cm 3.5cm 7.5cm 0cm},clip]{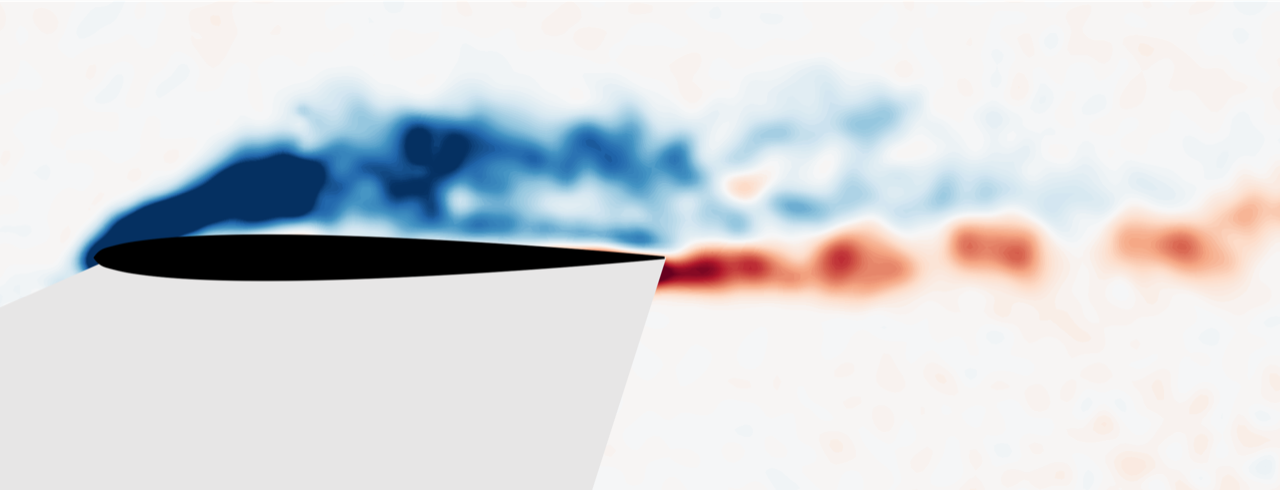}
	&\includegraphics[width=0.21\textwidth, trim={0.8cm 3.5cm 7.5cm 0cm},clip]{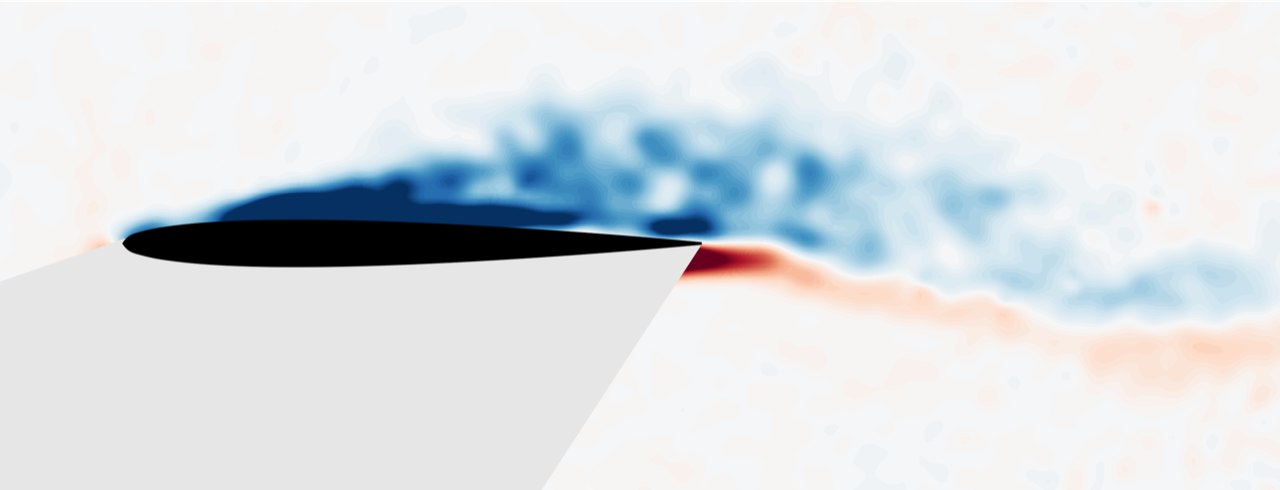}\\
	\bottomrule

\end{tabular}
	\includegraphics[width=0.65\textwidth]{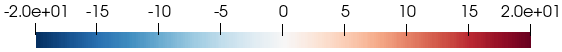}

		\caption{Comparison of experimental and CFD quarter span vorticity distributions for rectangular aspect-ratio-three
		wings oscillating in heave at $k=0.4$ and $h_0^*=0.5$.}
		\label{fig:vort_plots_validation}
\end{figure}

The validation case was that of the aspect-ratio-three wing oscillating at
amplitude $h_0^*=0.5$ and $k=0.4$. This amplitude is sufficient to cause the
 formation of leading edge vortices. A comparison of the lift coefficients
 obtained from the CFD and the experiment are shown in Fig.~\ref{fig:CFD_cl_validation}.
PIV data is compared at $t/T = [1/8, 3/16, 3/8, 1/2]$ - critical points that demonstrate
 the development of LEV at the quarter span on the wing. This is shown in
Fig.~\ref{fig:vort_plots_validation}.

The CFD and experiment are in good agreement for the lift coefficients.
The flow visualization from PIV data also agrees well with that from the CFD. Both methods
show the formation of the initialization of LEV formation between
$t/T=1/8$ and $t/T=3/16$. At $t/T=3/8$, the angle of the
leading edge shear layer matches, along with the approximate shape of
the LEV region. By $t/T = 1/2$, the methods both show that the vortex structure
has been convected downstream, and remains attached to the surface of the
airfoil.

\subsection{Lift and moment coefficient comparison}
\label{sec:whole_wing_forces}

\begin{figure}
\centering
    \subfigure[2D]{
	\label{fig:CL_comparison_arInf}
    	\includegraphics[width=0.4\textwidth]{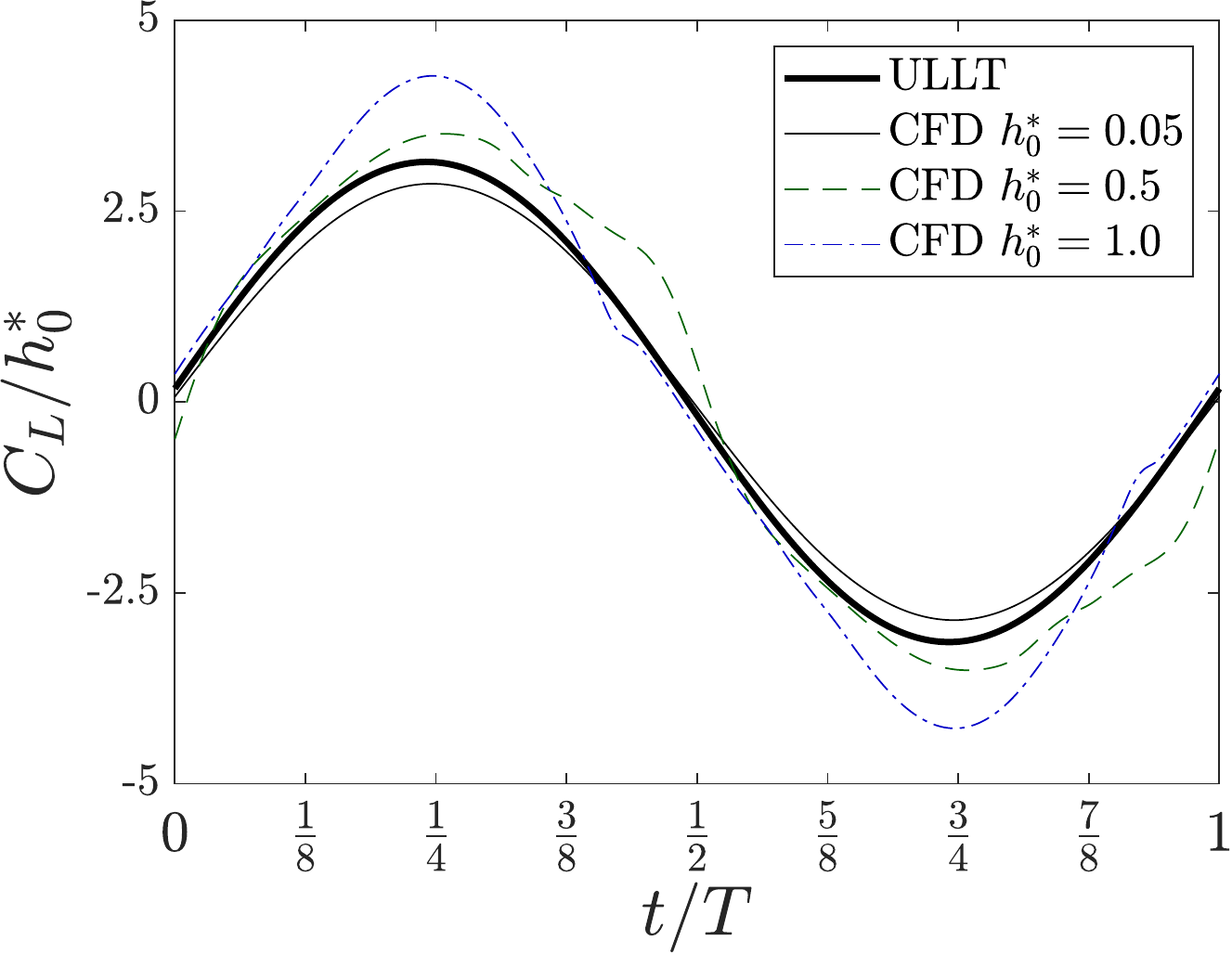}}
    \subfigure[Aspect ratio 6]{
	\label{fig:CL_comparison_ar6}
    	\includegraphics[width=0.4\textwidth]{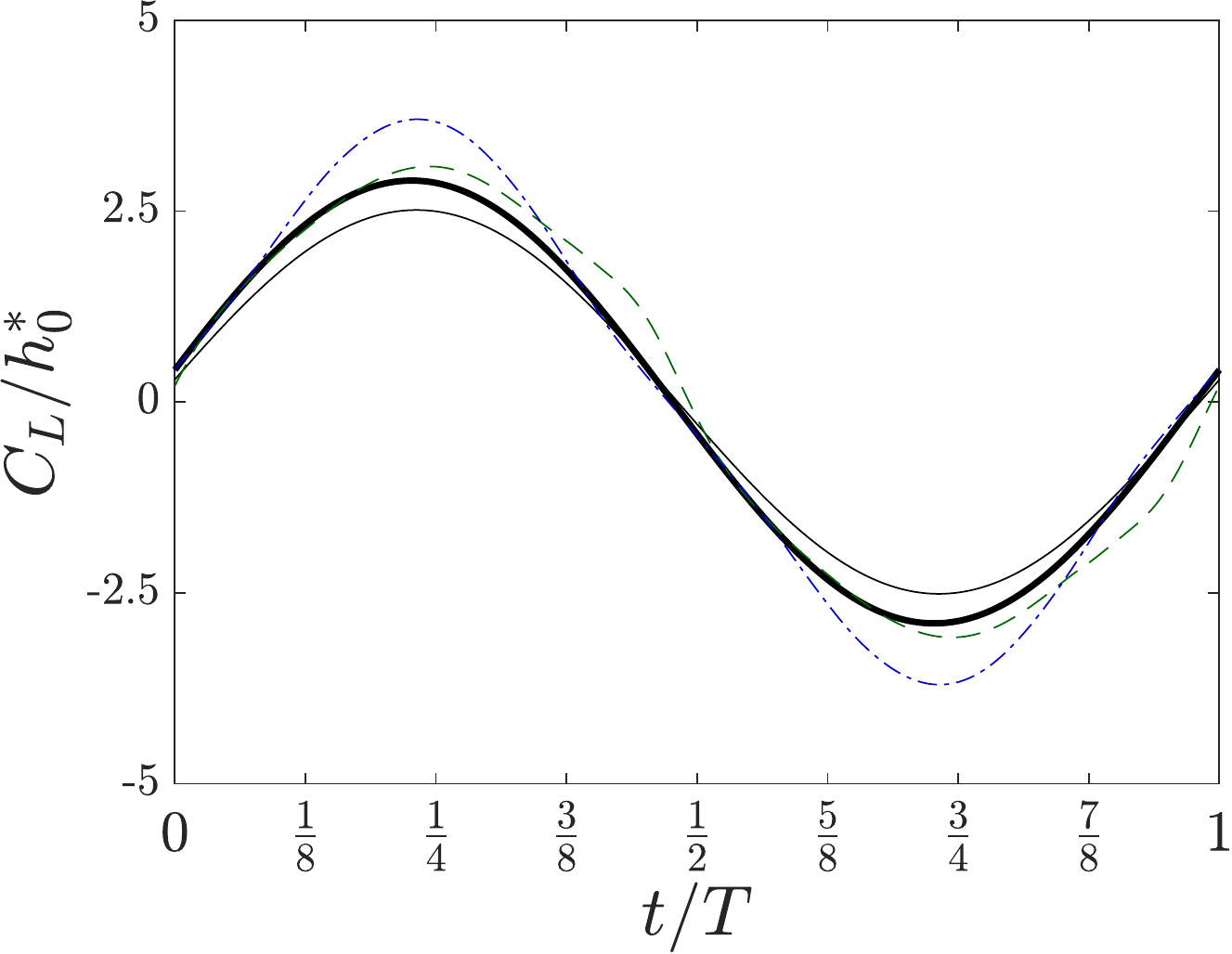}}

    \subfigure[Aspect ratio 3]{
	\label{fig:CL_comparison_ar3}
    	\includegraphics[width=0.4\textwidth]{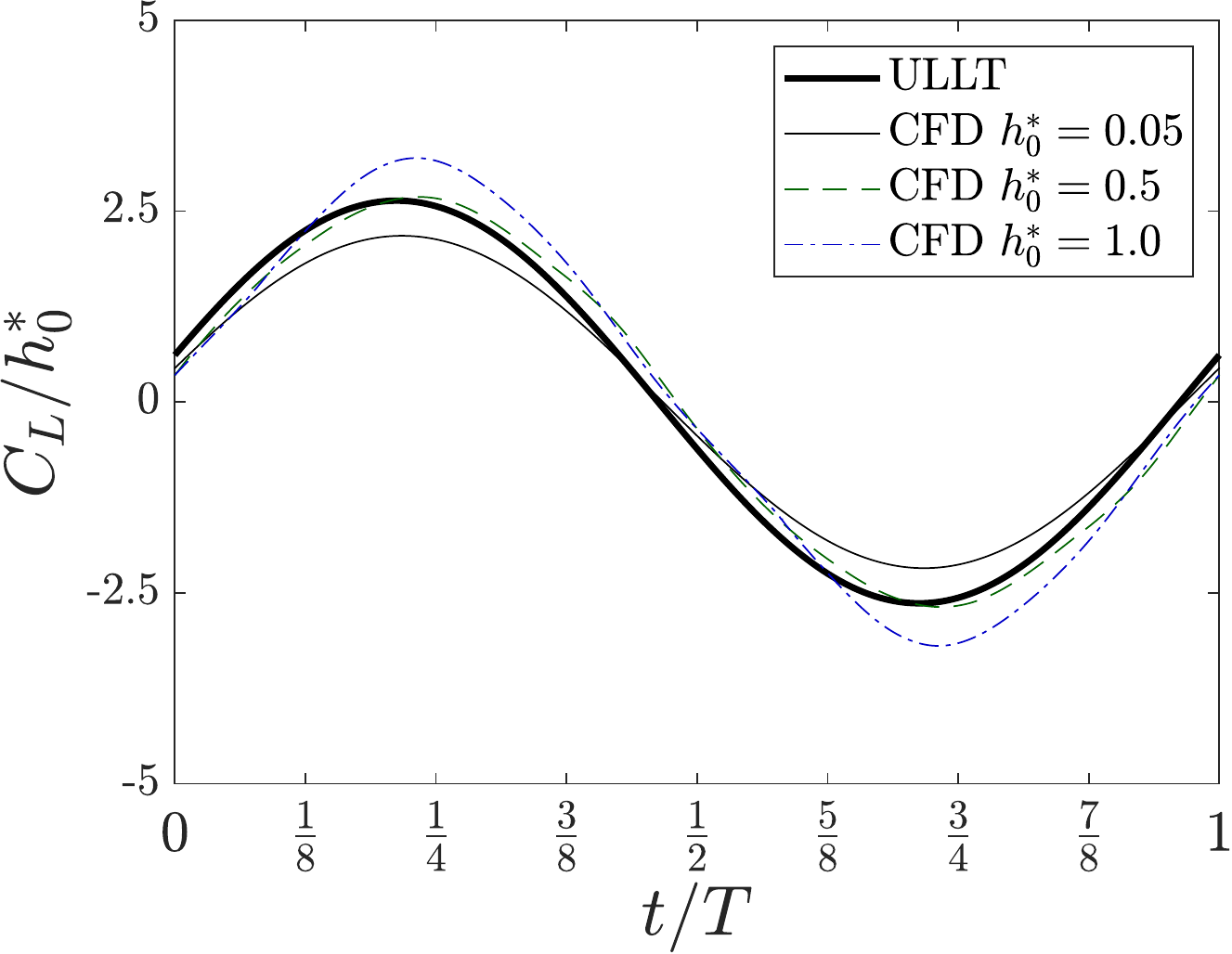}}
    \subfigure[Aspect ratio 1]{
	\label{fig:CL_comparison_ar1}
    	\includegraphics[width=0.4\textwidth]{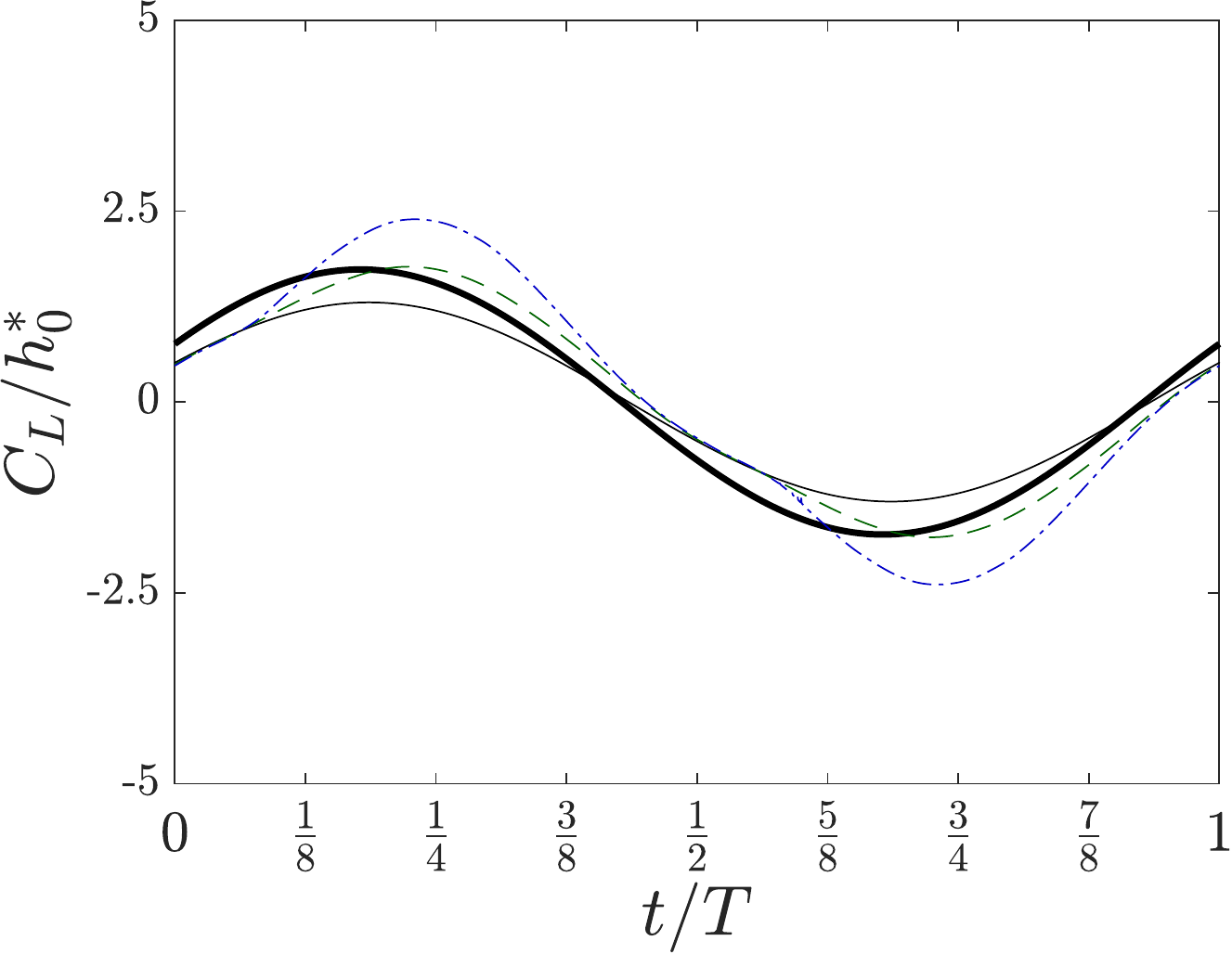}}

	\caption{Comparison of the predicted lift coefficients
	of ULLT and CFD for rectangular wings oscillating in heave
	at various amplitudes and aspect ratios at $k=0.4$.}
	\label{fig:CL_comparison}	
\end{figure}

Figure~\ref{fig:CL_comparison} shows a comparison of whole wing
lift coefficients for the cases listed in Table~\ref{tab:case_parameters}.
 The lift coefficient predicted by Sclavounos' ULLT 
varies linearly with oscillation amplitude, so lift coefficients have been
normalized by oscillation amplitude to allow for better comparison.

\begin{figure}
\centering
    \subfigure[LESP against time at wing center]{
	\label{fig:A0_vs_time}
    	\includegraphics[width=0.4\textwidth]{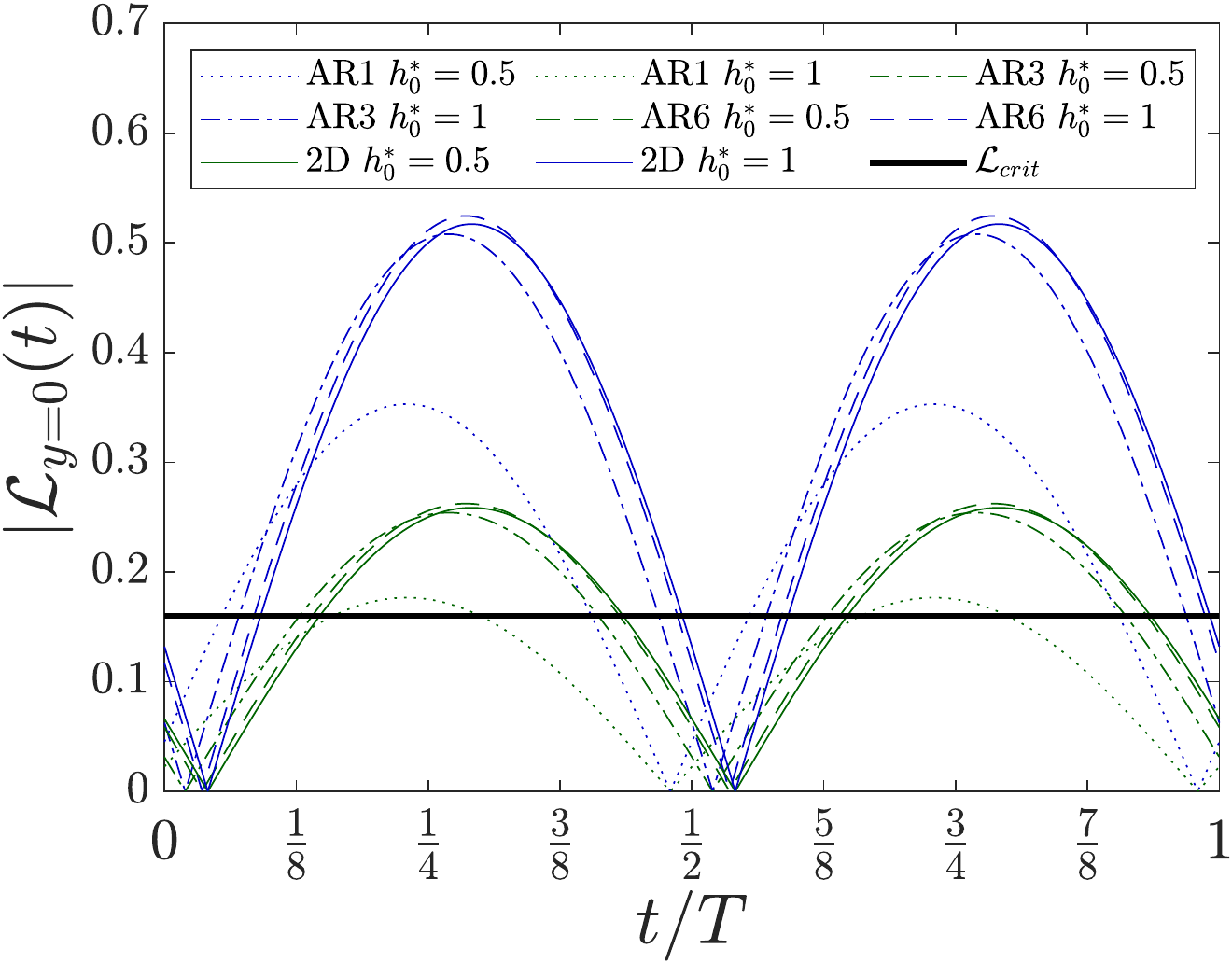}}
    \subfigure[LESP amplitude spanwise distribution]{
	\label{fig:A0_amplitude}
    	\includegraphics[width=0.4\textwidth]{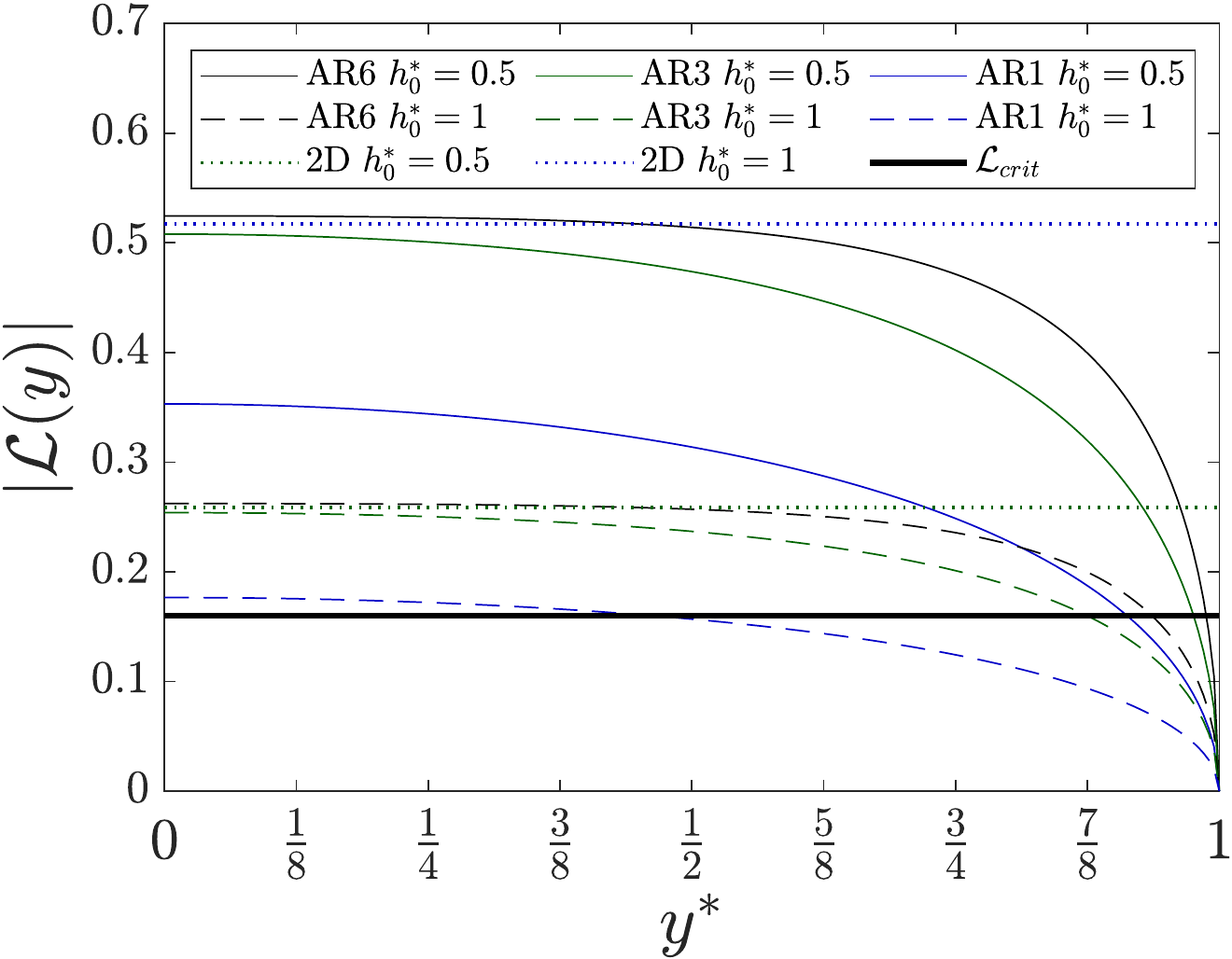}}

	\caption{Comparison of the predicted leading edge suction parameter
	of ULLT for rectangular wings oscillating in heave
	at various amplitudes and aspect ratios at $k=0.4$.}
	\label{fig:a0_wing}	
\end{figure}

The 2D problem, equivalent to an infinite aspect ratio wing, is shown in 
Fig.~\ref{fig:CL_comparison_arInf}. At low amplitude $h_0^*=0.05$, 
the CFD result is sinusoidal in form.
As oscillation amplitude increases to $h_0^*=0.5$ and then 
$h_0^*=1$, the peak lift increases super-linearly with respect to oscillation amplitude. 
This is a result of the leading-edge vortex being formed at these amplitudes. 
At the intermediate amplitude $h_0^*=0.5$,
the LEV is both shed later than in the $h_0^*=1$ case, and is slower to pinch off from the airfoil. As a consequence, the lift enhancement
comes later (visible in the humped shape of the $C_L$ curve).

The ULLT / Theodorsen predicted value of LESP from Eq.~\ref{eq:lesp_with_respect_to_span} is shown in Fig.~\ref{fig:a0_wing},
and can be used to predict this LEV shedding.
The time-varying value of the leading edge suction parameter at the wing center $\mathcal{L}_{y=0}$
is shown in Fig.~\ref{fig:A0_vs_time}.
Leading-edge vortex formation occurs when the critical value
of LESP $\mathcal{L}_{crit}$ is exceeded. 
The critical value $\mathcal{L}_{crit}$ was found by calibration to be 0.16
using the method described in Ramesh et al.~\cite{Ramesh2014}. 
This model predicts that LEV shedding will occur at all aspect ratios, for both $h_0^*=0.5$ and $h_0^*=1.0$.
The LESP values for the $h_0^*=0.05$  cases are not plotted since they are small
and do not approach the critical value of LESP.
Comparison between Fig.~\ref{fig:A0_vs_time}
and Fig.~\ref{fig:CL_comparison} does yield a relationship. However, the flow
around an airfoil for oscillating kinematics depends on historical LEV shedding. Since this ULLT
does not model LEVs, this means that it is unable to account  for this.
Consequently, it is more useful to focus on the amplitude of the oscillating LESP
value, as presented in Fig.~\ref{fig:A0_amplitude}. In 2D, the values of $\mathcal{L}$
obtained using Theodorsen's theory predict the LEV formation that occurs in the CFD results.
Additionally, the larger value of predicted LESP amplitude that occurs as the amplitude
of the kinematics increases anticipates the formation of a stronger LEV. Consequently,
the LEV has a greater impact on the force coefficients.

Unsteady lifting-line theory, equivalent to Theodorsen's theory at infinitely 
high aspect ratio, predicts both the phase and amplitude of the smallest
amplitude case well. The higher amplitude cases, $h_0^*=0.5$ and $h_0^*=1$,
have a non-sinusoidal $C_L$ waveform, unlike that predicted by the theory. 
Nevertheless, given the difference in flow field, explored in Sec.~\ref{sec:centre_wing_vorts} 
and Sec.~\ref{sec:force_distributions}, the accuracy
with which the CFD result can be predicted is reasonably good.

Figures~\ref{fig:CL_comparison_ar6},~\ref{fig:CL_comparison_ar3} and~\ref{fig:CL_comparison_ar1}
show the results of ULLT compared to CFD for decreasing aspect ratio.

As would be expected, the CFD shows that the lower aspect ratio wings
produce less lift.
This is due to the downwash induced by the trailing wing tip vortices
resulting in a lower effective plunge amplitude ($h_0(y) - F(y)$)
than found in the 2D case. This  effect cannot be
predicted by strip theory.
And at all aspect ratios the lift continues to vary super-linearly with
amplitude. Increasing the amplitude disproportionately increases
lift due to the effects of the leading edge vortex. This cannot be
predicted by the linearized ULLT used here.

Focusing on the result obtained by unsteady lifting-line theory for amplitude $h_0^*=0.05$,
it predicts both phase and amplitude 
well at aspect ratio 6, and the error isn't significantly larger 
than the difference between the Theodorsen prediction and the 2D CFD result. 
As aspect ratio decreases from \AR 6 to \AR 3 to \AR 1, 
the ULLT slightly over-predicts the amplitude of $C_L$, although
the sinusoidal waveform assumption and phase prediction remained good.
For the higher oscillation amplitudes, correctly predicted LEV formation occurs, 
and consequently the $C_L$ waveform from the CFD becomes 
non-sinusoidal. However, the normalized amplitude is still predicted
 reasonably well by the inviscid ULLT. 
For the cases with LEV shedding, the difference between CFD and ULLT because smaller
as $\AR$ decreases.

\begin{figure}
\centering
    \subfigure[2D]{
	\label{fig:CM_comparison_arInf}
    	\includegraphics[width=0.4\textwidth]{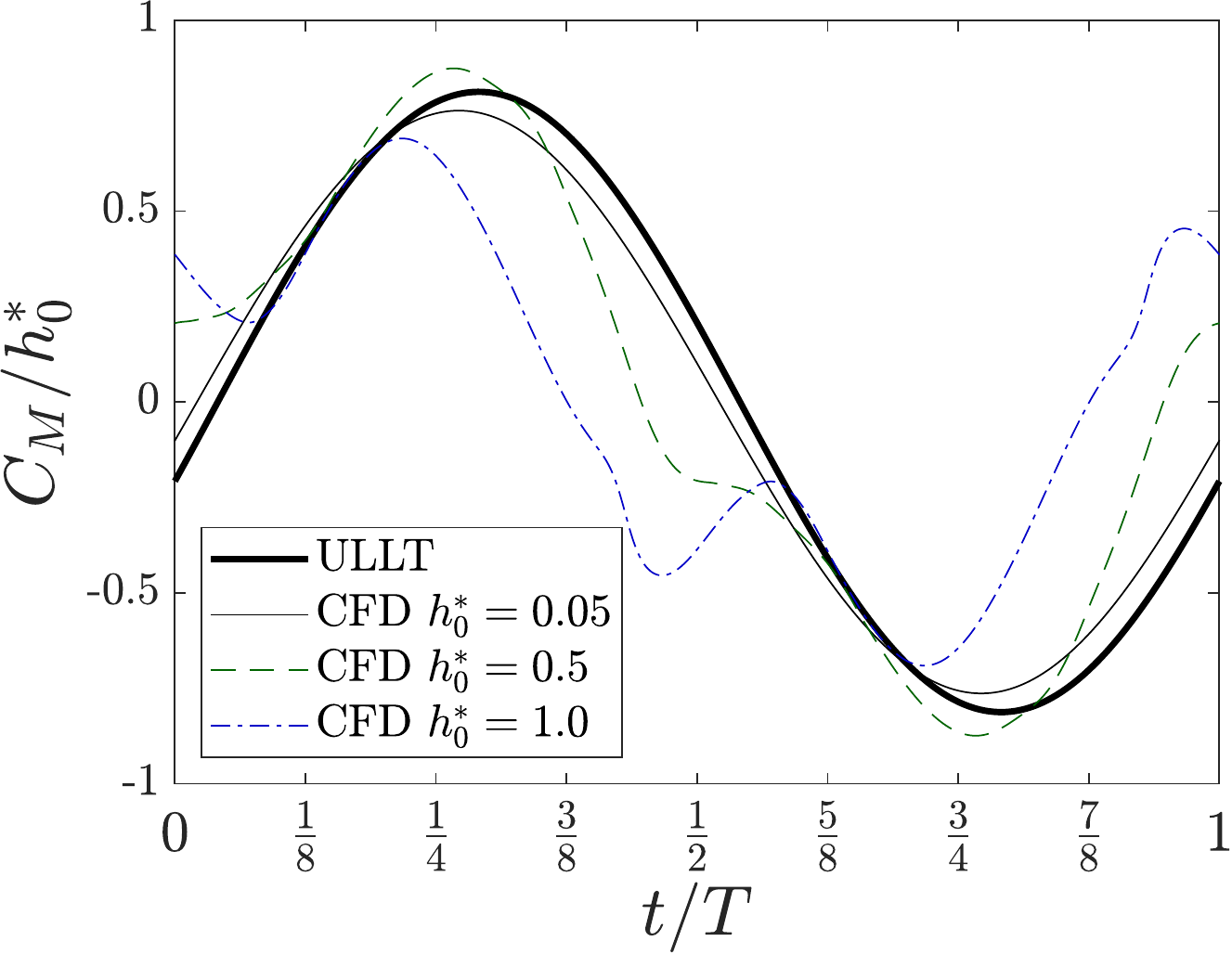}}
    \subfigure[Aspect ratio 6]{
	\label{fig:CM_comparison_ar6}
    	\includegraphics[width=0.4\textwidth]{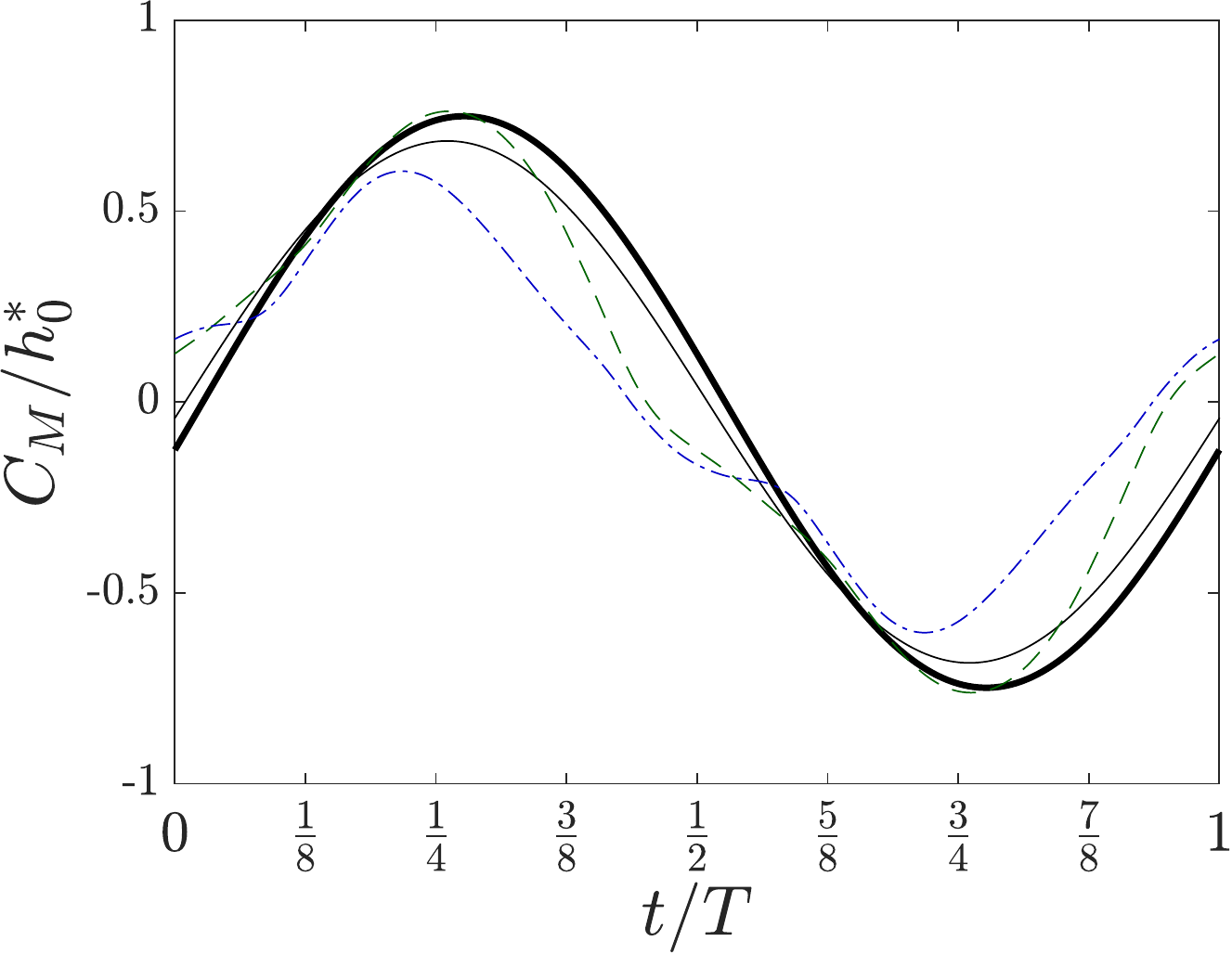}}

    \subfigure[Aspect ratio 3]{
	\label{fig:CM_comparison_ar3}
    	\includegraphics[width=0.4\textwidth]{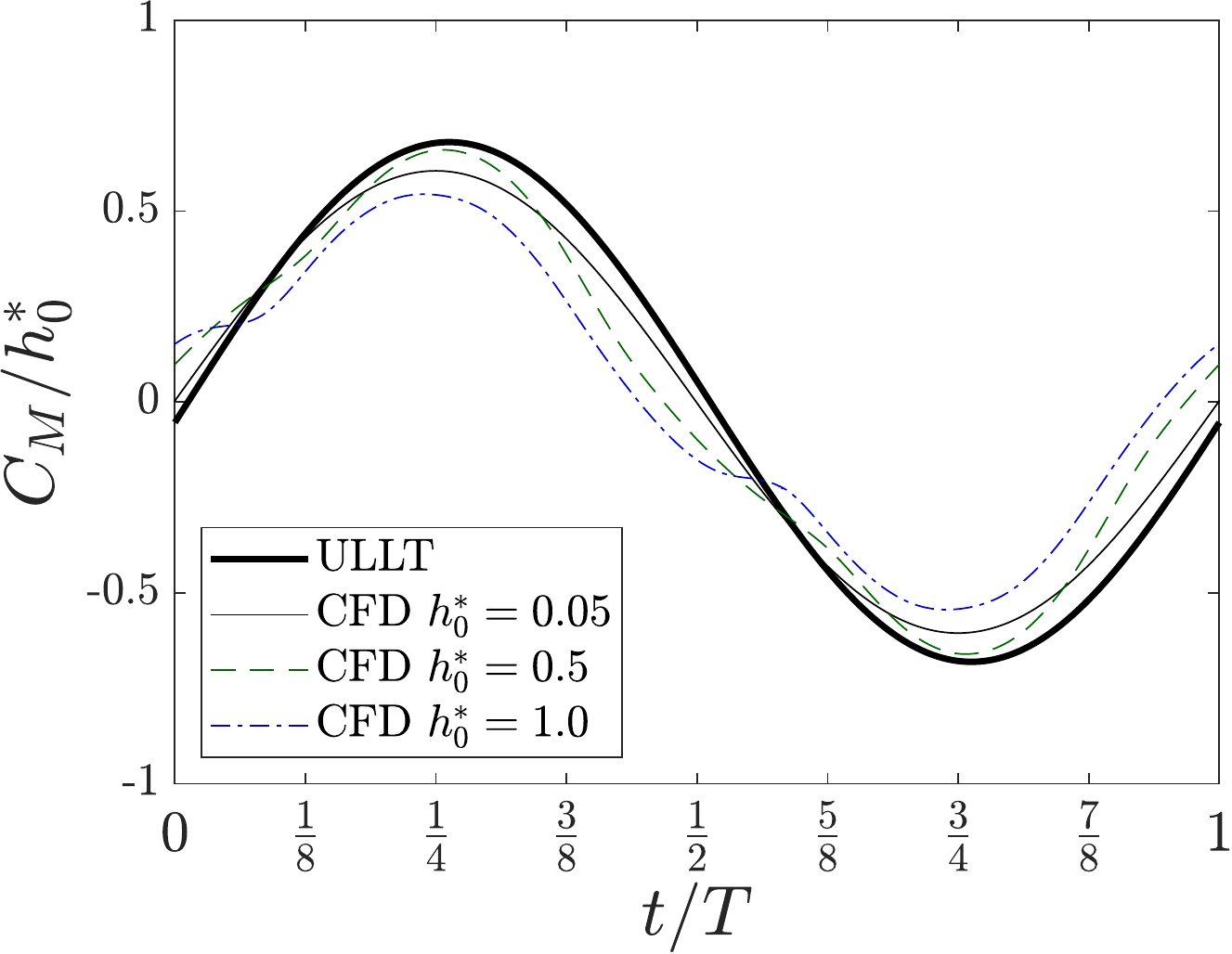}}
    \subfigure[Aspect ratio 1]{
	\label{fig:CM_comparison_ar1}
    	\includegraphics[width=0.4\textwidth]{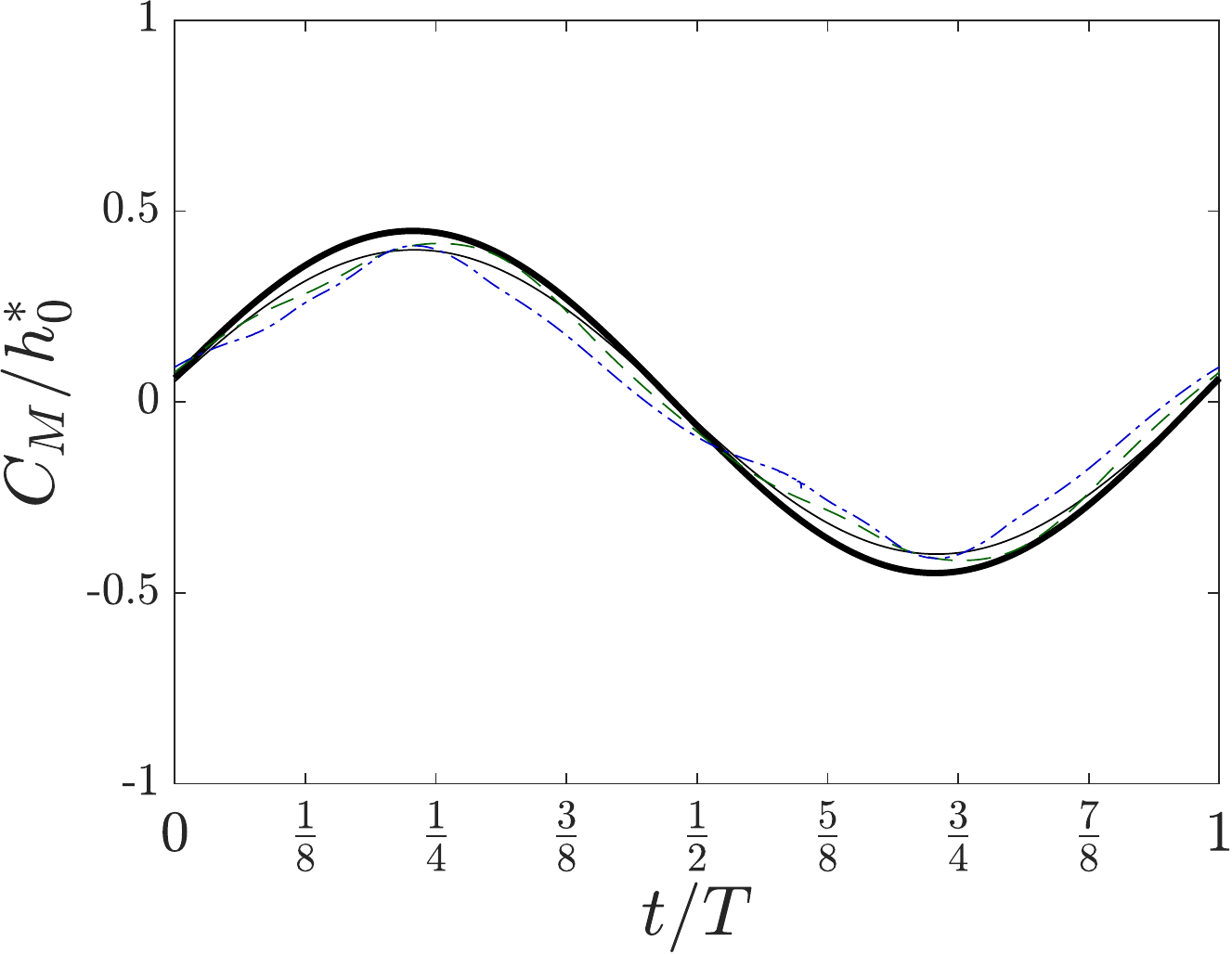}}

	\caption{Comparison of the predicted mid-chord moment coefficients
	of ULLT and CFD for rectangular wings oscillating in heave
	at various amplitudes and aspect ratios at $k=0.4$.}
	\label{fig:CM_comparison}	
\end{figure}

The whole wing moment coefficients are shown in Fig.~\ref{fig:CM_comparison}.
The 2D case $C_M$ curve is shown in Fig.~\ref{fig:CM_comparison_arInf}.
The low-amplitude heave kinematics results in a sinusoidal $C_M$ curve.
For the $h_0^*=0.5$ case
the normalized $C_M$ is initially similar to that of the low-amplitude
casen. As the LEV develops and sheds, the
$C_M$ initially increases and then drops markedly, resulting in a non-sinusoidal wave. The
$C_M$ briefly returns to the same normalized amplitude as the low amplitude
case before the another LEV forms during the second half of the cycle.
The $h_0^*=1$ case shows the same trends but with LEV initiation happening earlier and
the subsequent loss of $C_M$ being greater.
As the aspect ratio decreases, the high amplitude results obtained from the CFD
tend to the low amplitude $h_0^*=0.05$ case. Both the initial LEV-produced
increase and the subsequent reduction in $C_M$ during LEV shedding are diminished by finite wing effects.

At all aspect ratios the ULLT predicts the $C_M$ results of the low
amplitude $h_0^*=0.05$ cases with good accuracy in term 
of both amplitude and phase.
The ULLT cannot predict the non-sinusoidal $C_M$ curves resulting from
the large amplitude cases at high aspect ratio, although the amplitude
and phase are approximately correct. The dwindling effects of the LEV
on $C_M$ as aspect ratio decreases results in the ULLT providing a better
prediction at lower aspect ratio.

For both the lift and moment results, as aspect ratio decreases
the impact of the LEV decreases and the waveform becomes more sinusoidal.
Whilst ULLT cannot model LEVs, the LESP obtained from ULLT can be used to predict LEV formation. 
It predicts that
larger amplitudes will result in a larger value of LESP. This is reflected in
the super-linear relationship between $C_L$ and amplitude at all aspect ratios.
For $C_M$, increased values of LESP correspond to a larger deviation from the
sinusoidal waveform produced by low amplitude results. As aspect ratio
decreases this deviation from sinusoidal reduces, despite the fact
that LEV formation is still predicted by ULLT. In the next section (Sec.~\ref{sec:centre_wing_vorts}),
the form of the LEV obtained from CFD is examined more closely.

\subsection{Wing center vorticity distribution}
\label{sec:centre_wing_vorts}

\begin{figure}
\centering
\begin{tabular}{ >{\centering\arraybackslash}m{0.065\textwidth}  >{\centering\arraybackslash}m{0.15\textwidth}   >{\centering\arraybackslash}m{0.15\textwidth} >{\centering\arraybackslash}m{0.15\textwidth} >{\centering\arraybackslash}m{0.15\textwidth}  >{\centering\arraybackslash}m{0.15\textwidth}}
	\toprule
	& $t/T = 1/16$ & $t/T = 1/8$ & $t/T = 3/16 $ & $t/T =3/8$ & $t/T =1/2$ \\
	\midrule	
	2D $h_0^*=1$
	&\includegraphics[width=0.16\textwidth]{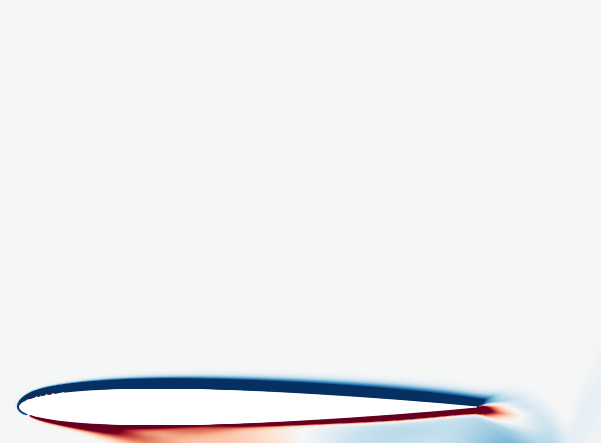}
	&\includegraphics[width=0.16\textwidth]{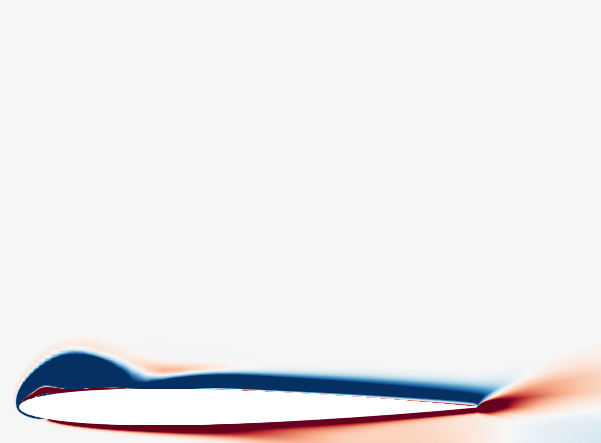}
	&\includegraphics[width=0.16\textwidth]{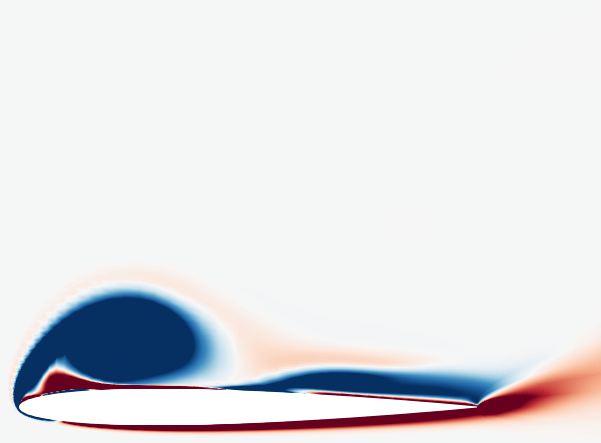}
	&\includegraphics[width=0.16\textwidth]{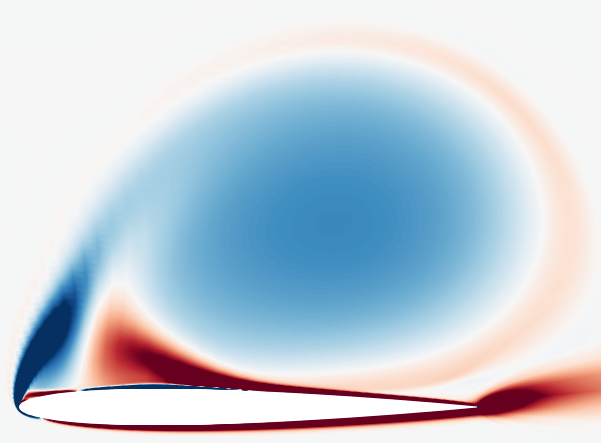}
	&\includegraphics[width=0.16\textwidth]{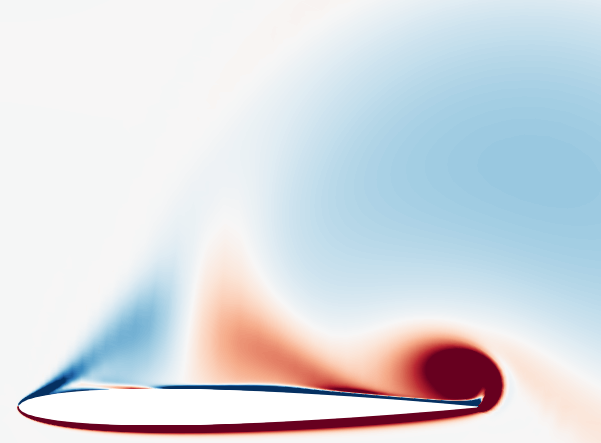}\\

	\AR6 $h_0^*=1$
	&\includegraphics[width=0.16\textwidth]{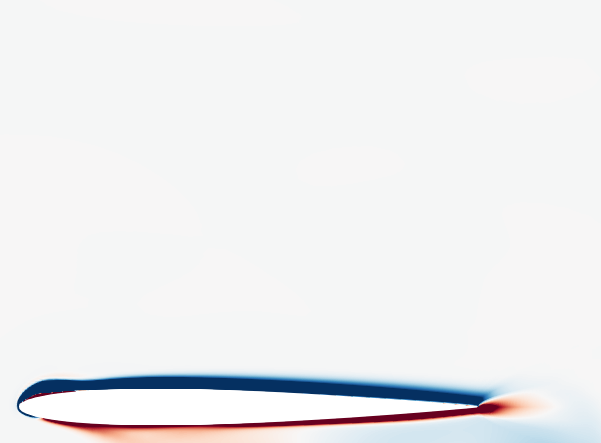}
	&\includegraphics[width=0.16\textwidth]{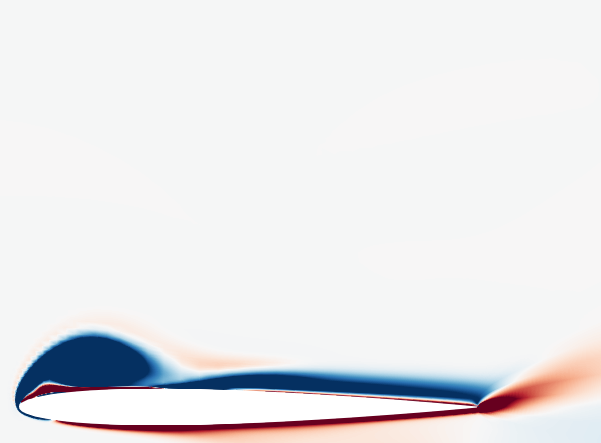}
	&\includegraphics[width=0.16\textwidth]{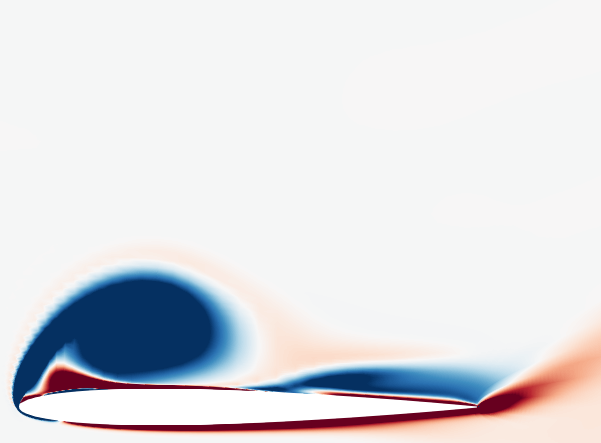}
	&\includegraphics[width=0.16\textwidth]{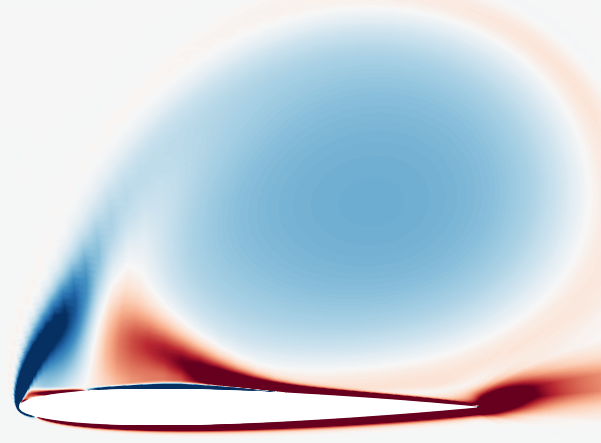}
	&\includegraphics[width=0.16\textwidth]{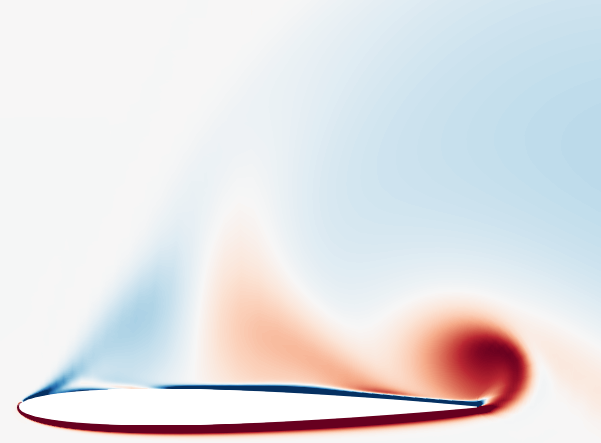}\\
	
	\AR3 $h_0^*=1$
	&\includegraphics[width=0.16\textwidth]{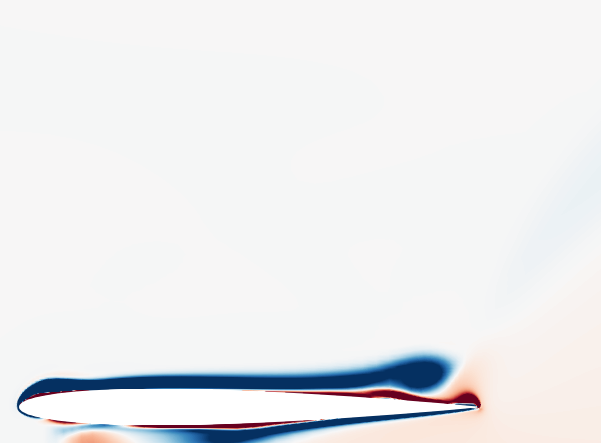}
	&\includegraphics[width=0.16\textwidth]{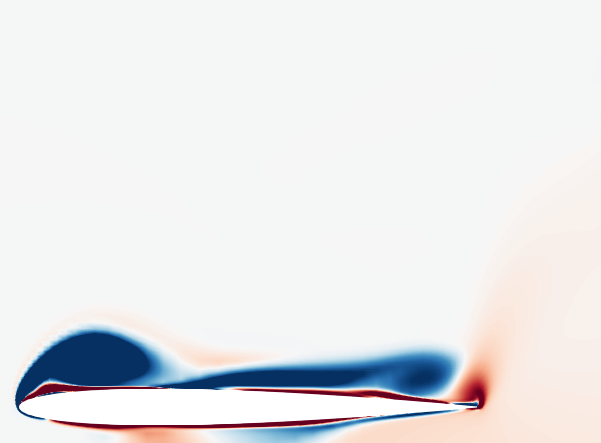}
	&\includegraphics[width=0.16\textwidth]{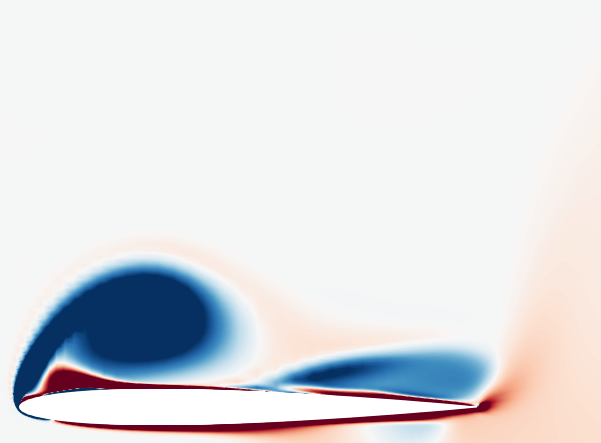}
	&\includegraphics[width=0.16\textwidth]{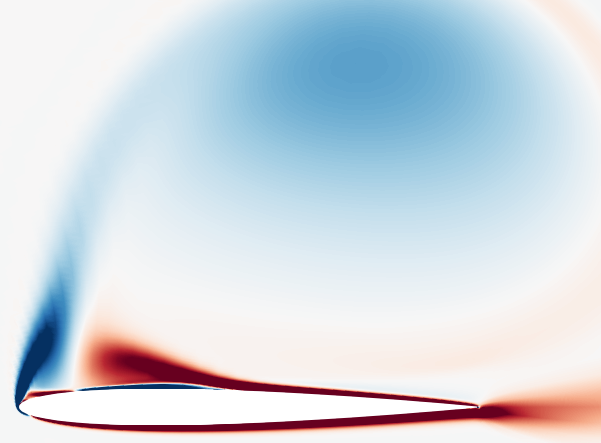}
	&\includegraphics[width=0.16\textwidth]{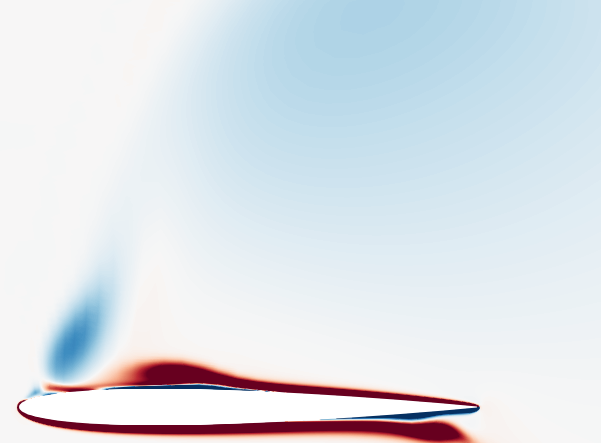}\\

	\AR1 $h_0^*=1$
	&\includegraphics[width=0.16\textwidth]{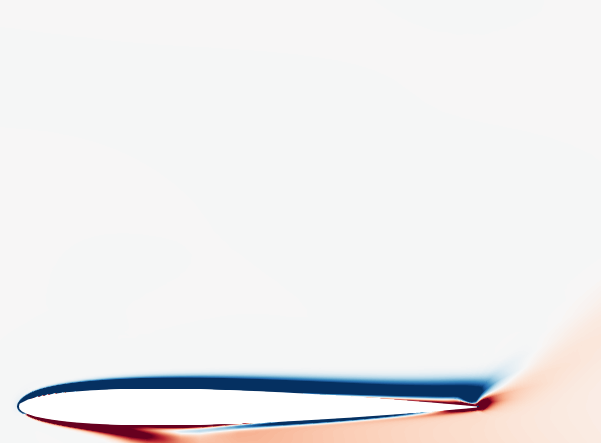}
	&\includegraphics[width=0.16\textwidth]{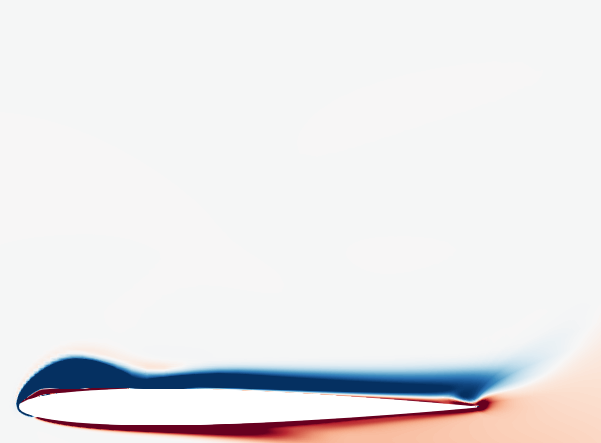}
	&\includegraphics[width=0.16\textwidth]{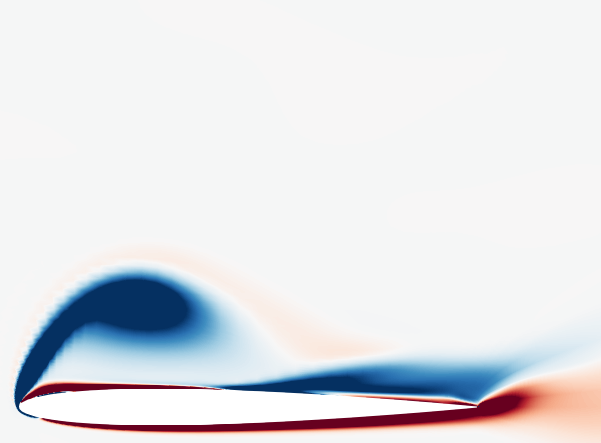}
	&\includegraphics[width=0.16\textwidth]{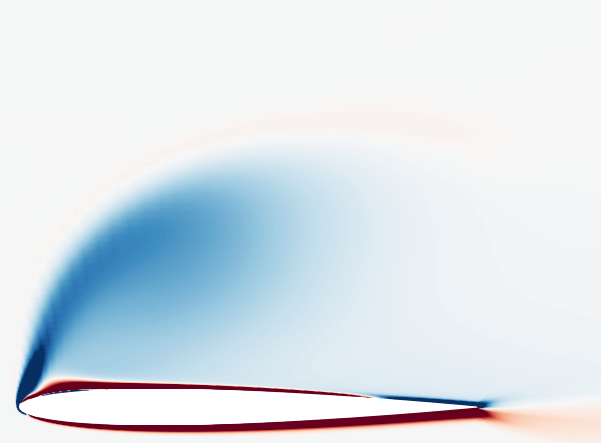}
	&\includegraphics[width=0.16\textwidth]{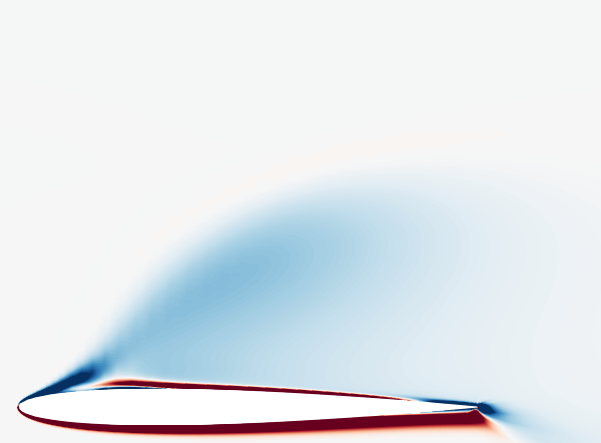}\\
	
	\midrule 
	2D $h_0^*=\frac{1}{2}$
	&\includegraphics[width=0.16\textwidth, trim={0 0 0 8cm},clip]{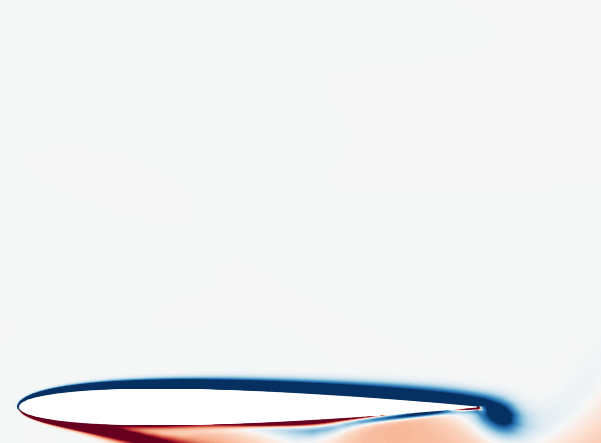}
	&\includegraphics[width=0.16\textwidth, trim={0 0 0 8cm},clip]{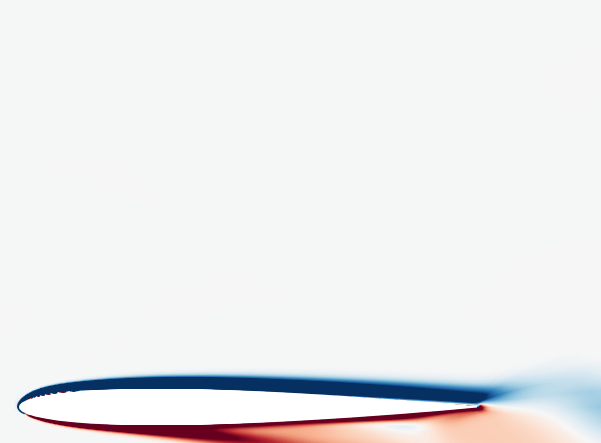}
	&\includegraphics[width=0.16\textwidth, trim={0 0 0 8cm},clip]{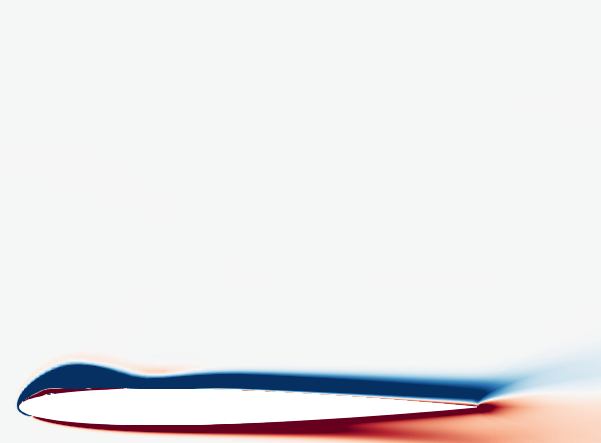}
	&\includegraphics[width=0.16\textwidth, trim={0 0 0 8cm},clip]{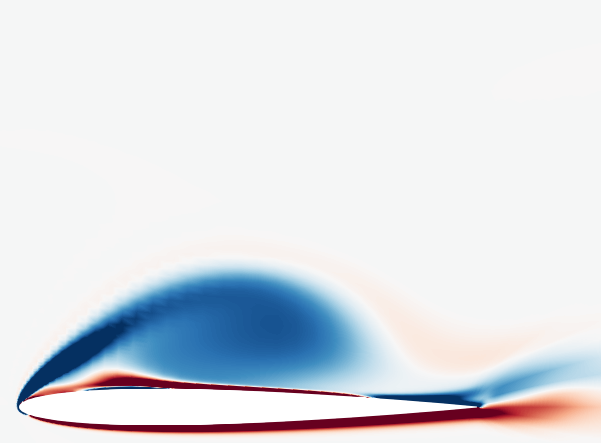}
	&\includegraphics[width=0.16\textwidth, trim={0 0 0 8cm},clip]{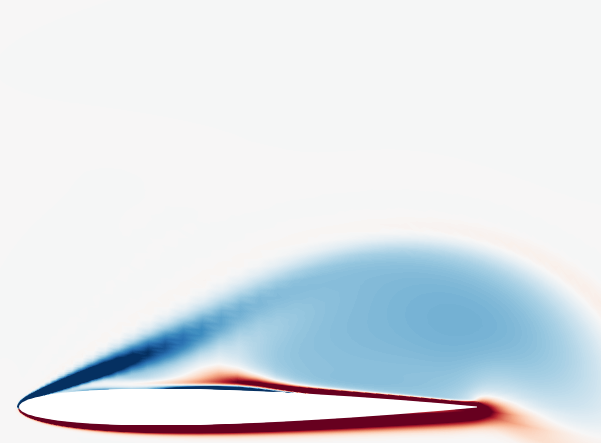}\\

	\AR6 $h_0^*=\frac{1}{2}$
	&\includegraphics[width=0.16\textwidth, trim={0 0 0 8cm},clip]{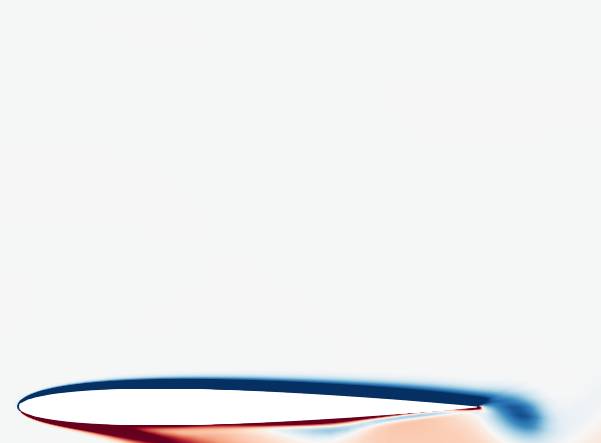}
	&\includegraphics[width=0.16\textwidth, trim={0 0 0 8cm},clip]{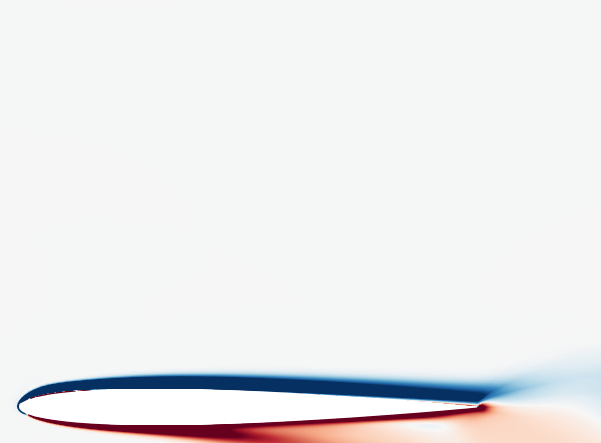}
	&\includegraphics[width=0.16\textwidth, trim={0 0 0 8cm},clip]{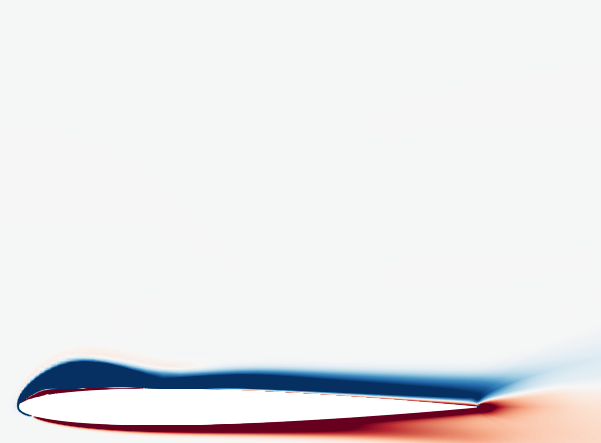}
	&\includegraphics[width=0.16\textwidth, trim={0 0 0 8cm},clip]{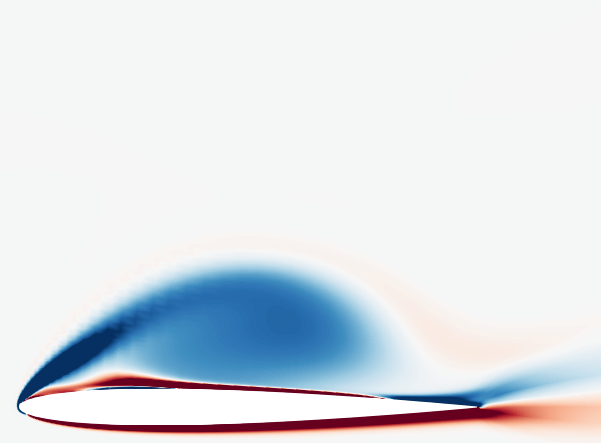}
	&\includegraphics[width=0.16\textwidth, trim={0 0 0 8cm},clip]{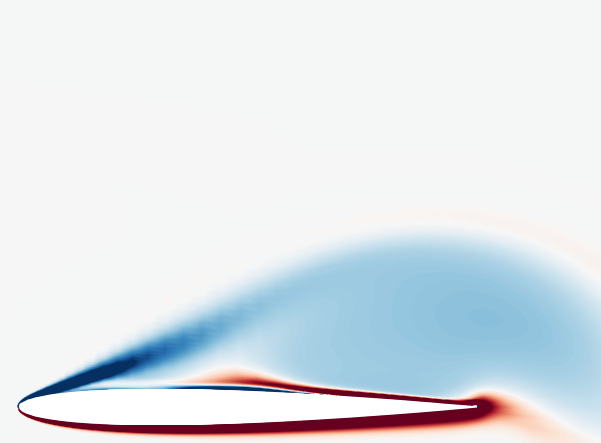}\\
	
	\AR3  $h_0^*=\frac{1}{2}$
	&\includegraphics[width=0.16\textwidth, trim={0 0 0 8cm},clip]{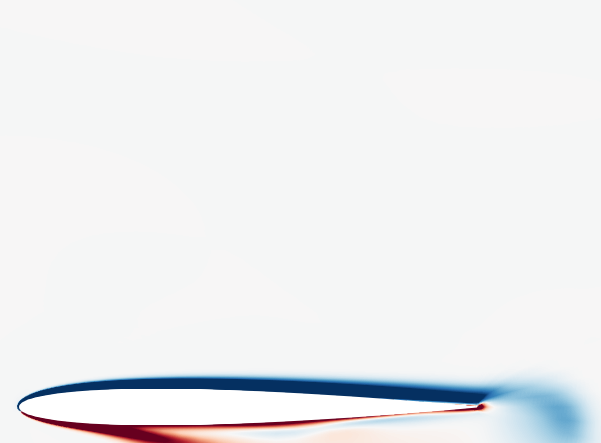}
	&\includegraphics[width=0.16\textwidth, trim={0 0 0 8cm},clip]{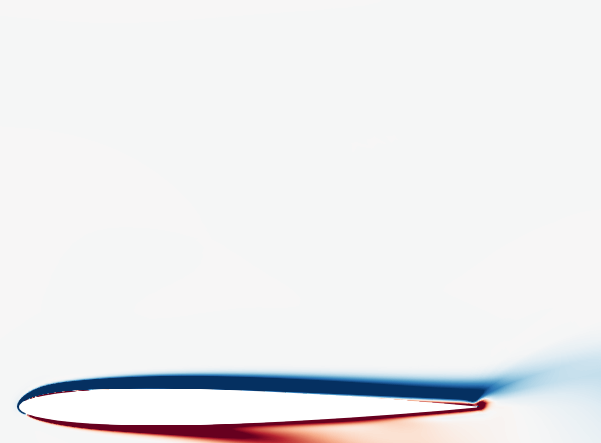}
	&\includegraphics[width=0.16\textwidth, trim={0 0 0 8cm},clip]{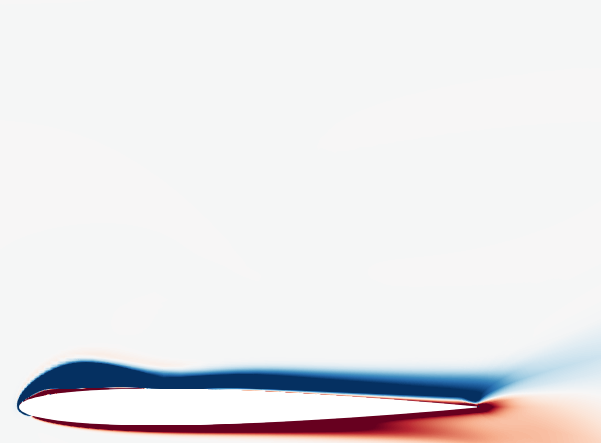}
	&\includegraphics[width=0.16\textwidth, trim={0 0 0 8cm},clip]{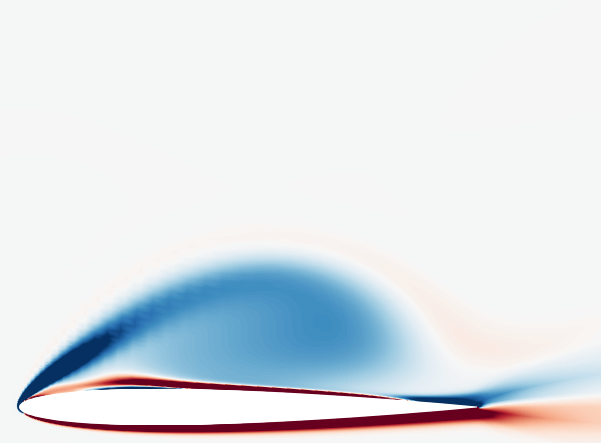}
	&\includegraphics[width=0.16\textwidth, trim={0 0 0 8cm},clip]{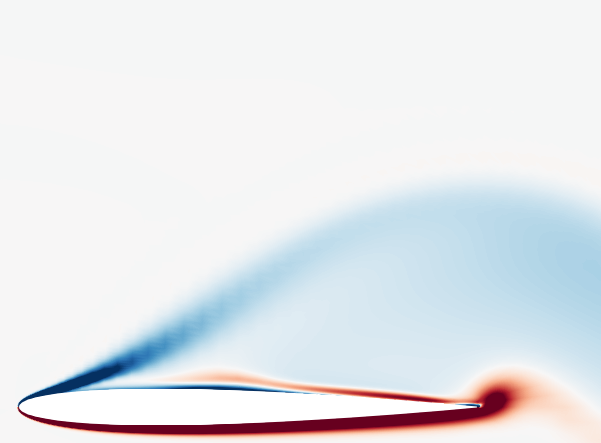}\\

	\AR1  $h_0^*=\frac{1}{2}$
	&\includegraphics[width=0.16\textwidth, trim={0 0 0 8cm},clip]{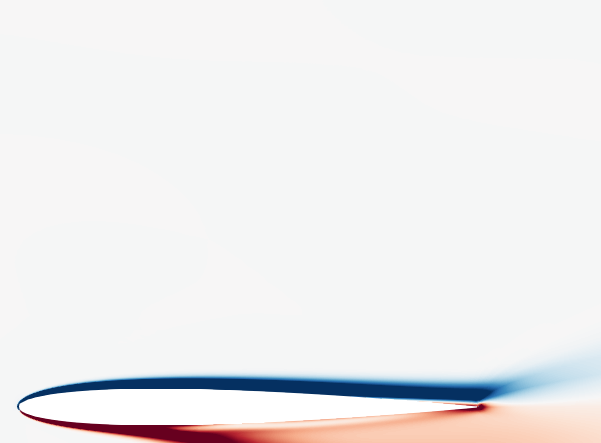}
	&\includegraphics[width=0.16\textwidth, trim={0 0 0 8cm},clip]{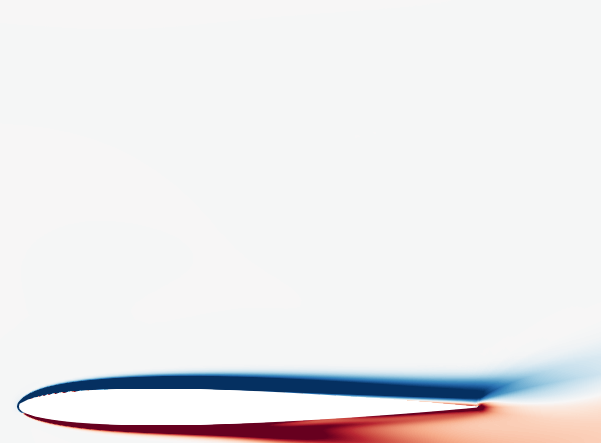}
	&\includegraphics[width=0.16\textwidth, trim={0 0 0 8cm},clip]{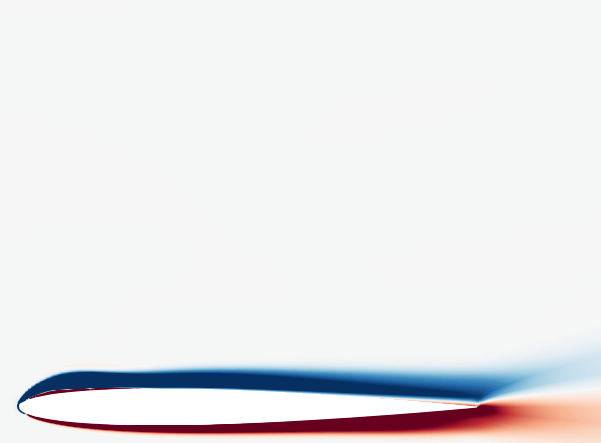}
	&\includegraphics[width=0.16\textwidth, trim={0 0 0 8cm},clip]{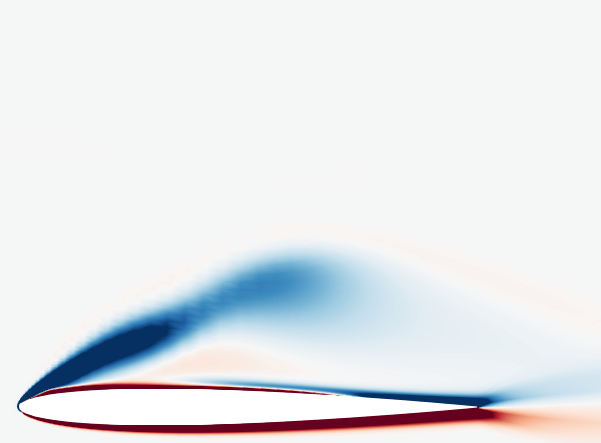}
	&\includegraphics[width=0.16\textwidth, trim={0 0 0 8cm},clip]{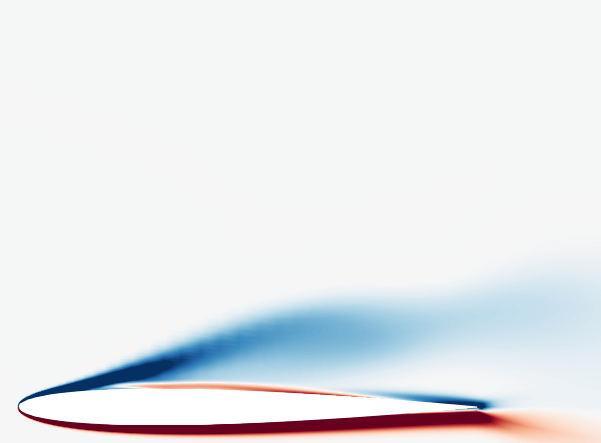}\\
	
	\bottomrule
\end{tabular}
	\includegraphics[width=0.65\textwidth]{figs/cfd_vort_20_hce_bar}
		\caption{Wing center spanwise vorticity distributions for rectangular
		wings oscillating in heave at $k=0.4$. The in plane vorticity is	
		normalized as $\omega^* = \omega_y \overline{c} / U$}
		\label{fig:wing_centre_vort_plots}
\end{figure}

The spanwise component of vorticity at the
center of the wings has been
plotted in Fig.~\ref{fig:wing_centre_vort_plots} for 
amplitudes $h_0^*=0.5$ and $h_0^*=1$. The
low amplitude cases with $h_0^*=0.05$ is not shown since it 
did not produces leading edge vortices, as predicted by ULLT.

For both amplitudes, the aspect ratio 6 case provides results similar to 2D. For
the $h_0^*=0.5$ case, at $t/T = 1/16$ and $t/T = 1/8$ the boundary layer on the wing's upper surface is attached.
As the plunge stroke continues the plunge velocity increases and consequentially
a greater leading edge suction required in order for flow to remain attached around he leading edge.
At $t/T=3/16$, an attached leading-edge vortex has formed. As the downstroke continues,
the leading edge vortex becomes larger, and is convected downstream. 
At $t/T=3/8$ the downstroke velocity is reducing, but the main vortex structure
remains in close proximity to the airfoil surface. 
The shear layer feeding the LEV becomes a prominent feature. 
At $t/T=1/2$, the downstroke has finished. The LEV
has now been convected to the trailing edge and has vorticity has spread out
due to viscous effects. The shear layer from the leading edge remains attached,
but has weakened.

At the higher amplitude $h_0^*=1$ the process is initially similar, but starts earlier
due to the increased velocity of the downstroke. An attached LEV has formed
by $t/T=1/8$. By $t/T=3/16$, the LEV has grown. A large counter rotating
region of vorticity has been drawn under the LEV. 
This region has grown significantly by $t/T=3/8$ and has separated
the LEV from the wing surface.The LEV is now larger and more
diffused, but also some distance from the wing surface.
By $t/T=1/2$, the LEV has completely detached from the shear layer
that initially fed it. It induces a counter vortex at the
trailing edge.

3D effects allow the LEV to remain attached to the leading edge and dissipate,
instead of detaching and being carried downstream. This effect is most
visible for the $h_0^*=1$, \AR1 case and results in simpler
$C_L$ and $C_M$ curves. For higher aspect ratio wings the LEV
at the wing center is similar to that of the 2D problem.
However, near to the wing tip the induced downwash
decreases the effective angle of attack
leading to a smaller vortex (see Fig.~\ref{fig:A0_amplitude}).

In the next section (Sec.~\ref{sec:force_distributions}), the spanwise 3D vortex structures and force distributions
are examined, and the claim that the impact of the LEV is
reduced near the wing-tips substantiated. 

\subsection{Spanwise force distributions}
\label{sec:force_distributions}

\begin{figure}
\centering
    \subfigure[Aspect ratio 6 $C_l$]{
	\label{fig:Cl_la_ar6}
    	\includegraphics[width=0.4\textwidth]{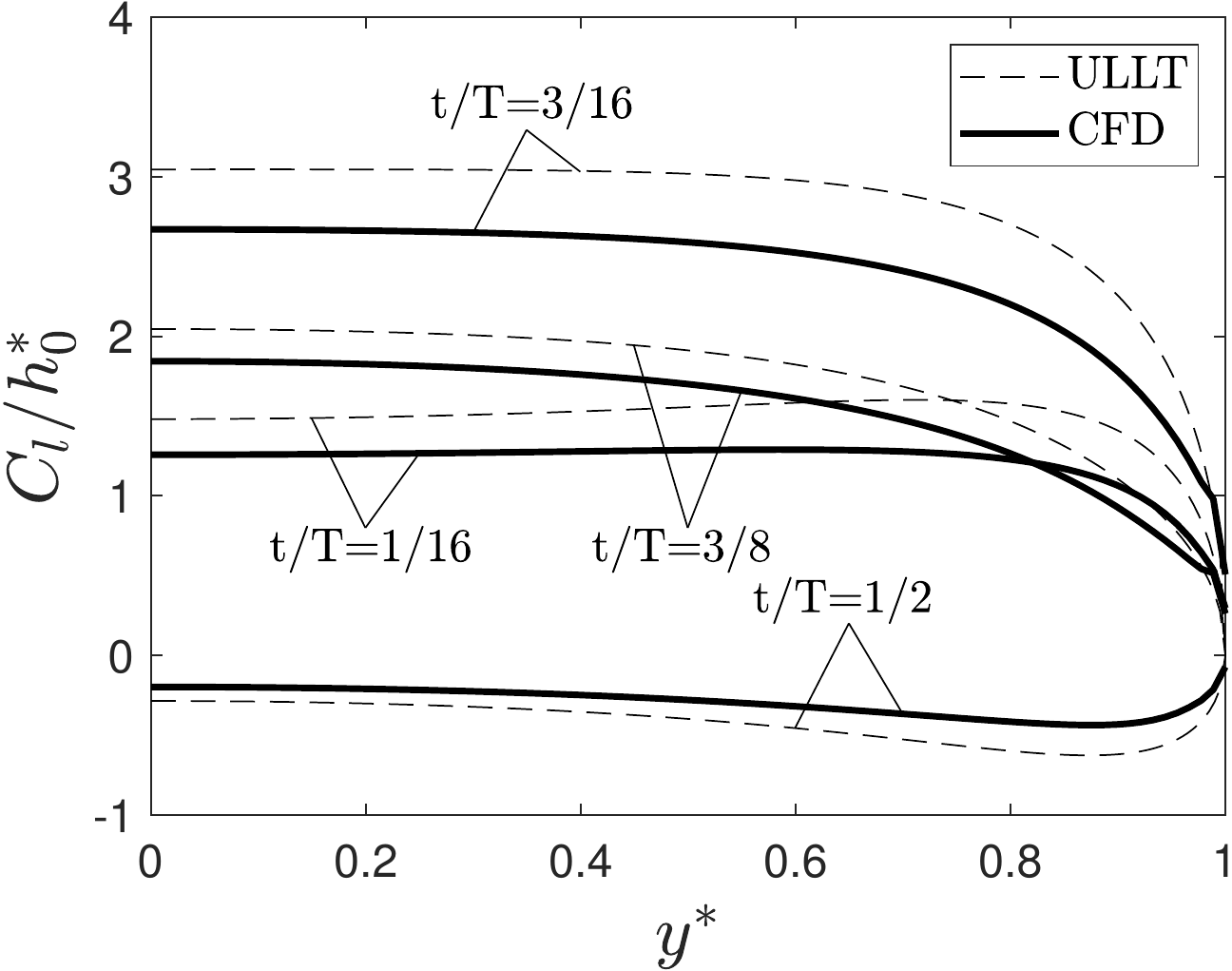}}
    \subfigure[Aspect ratio 1 $C_l$]{
	\label{fig:Cl_la_ar1}
    	\includegraphics[width=0.4\textwidth]{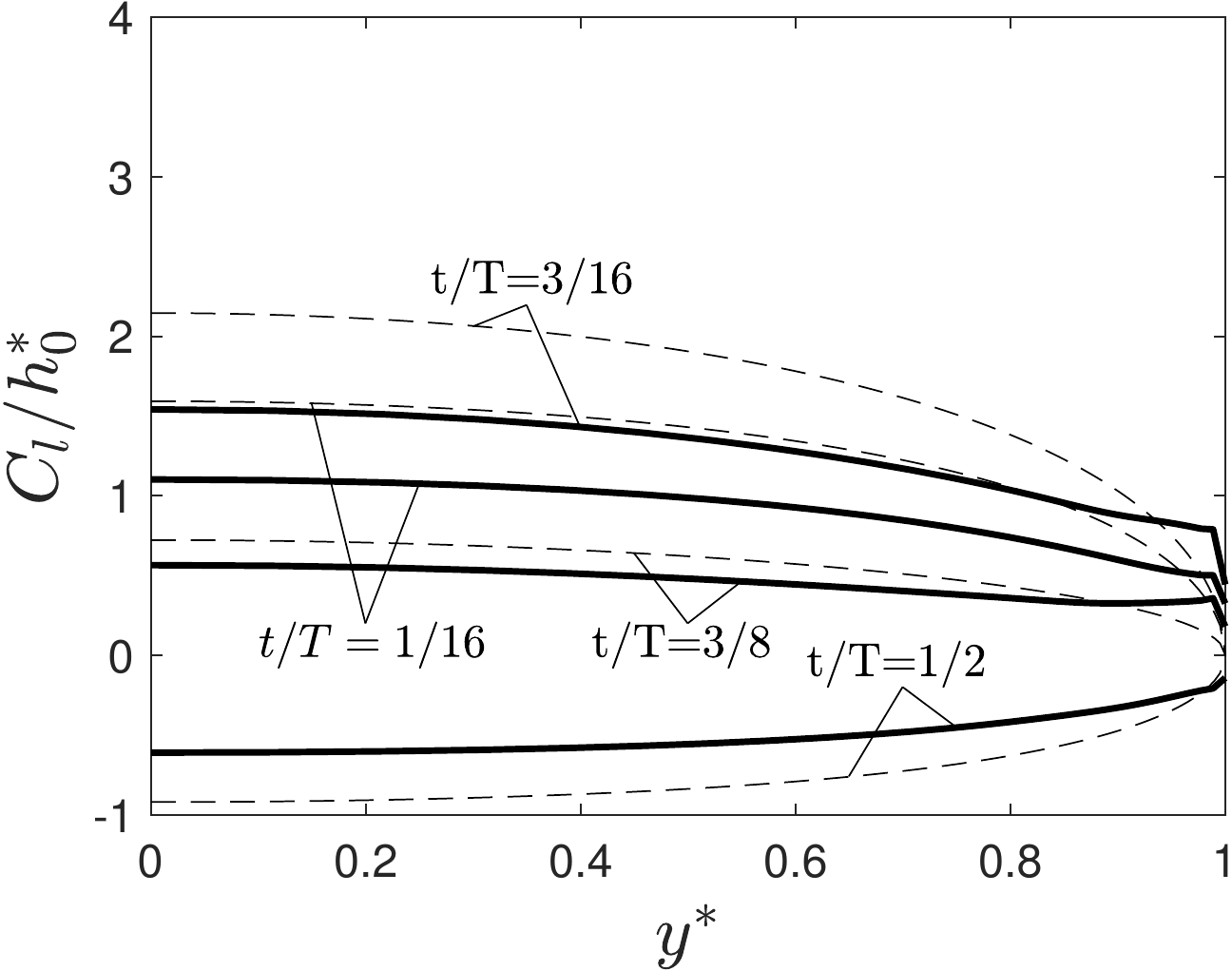}}

    \subfigure[Aspect ratio 6 $C_m$]{
	\label{fig:Cm_la_ar6}
    	\includegraphics[width=0.4\textwidth]{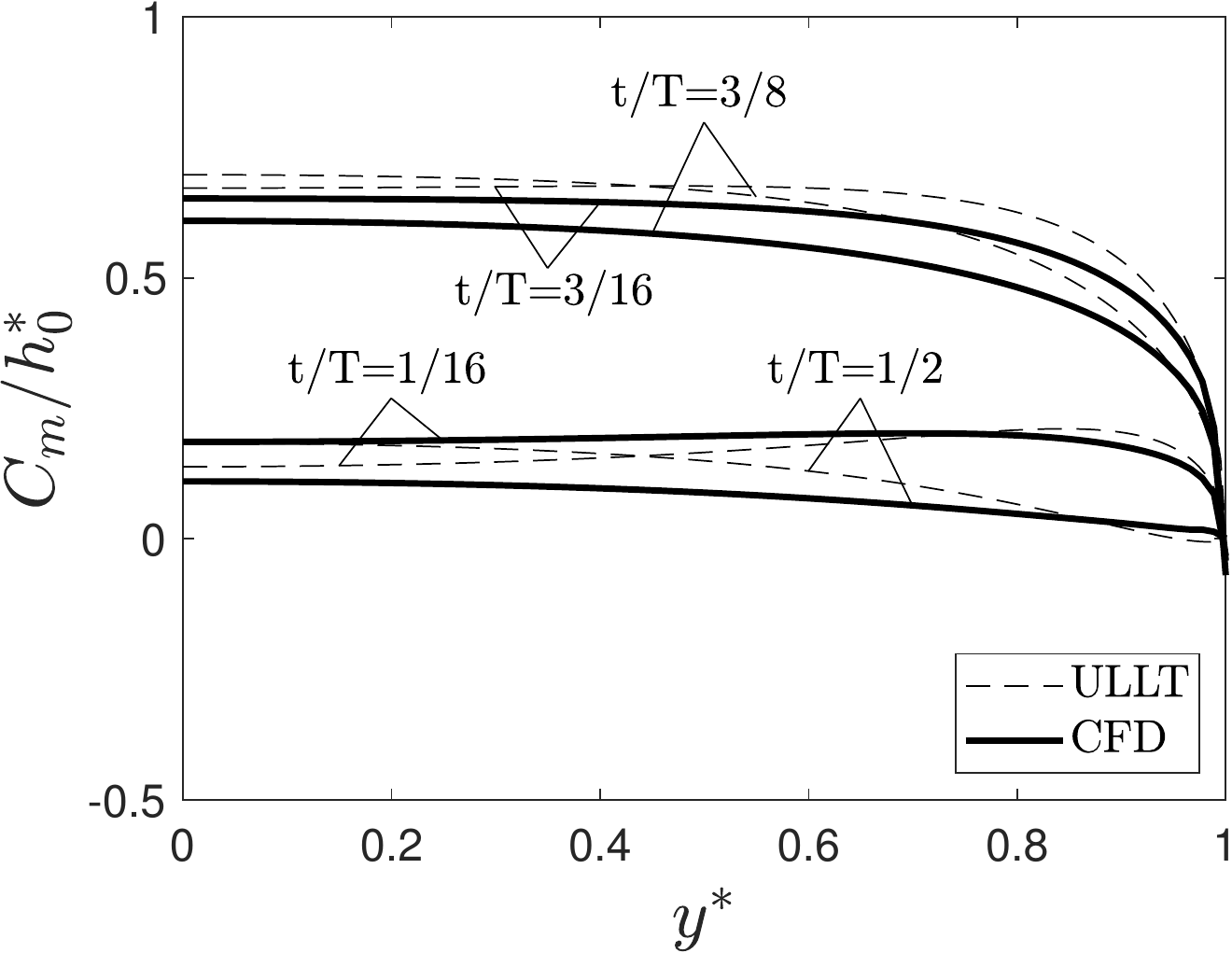}}
    \subfigure[Aspect ratio 1 $C_m$]{
	\label{fig:Cm_la_ar1}
    	\includegraphics[width=0.4\textwidth]{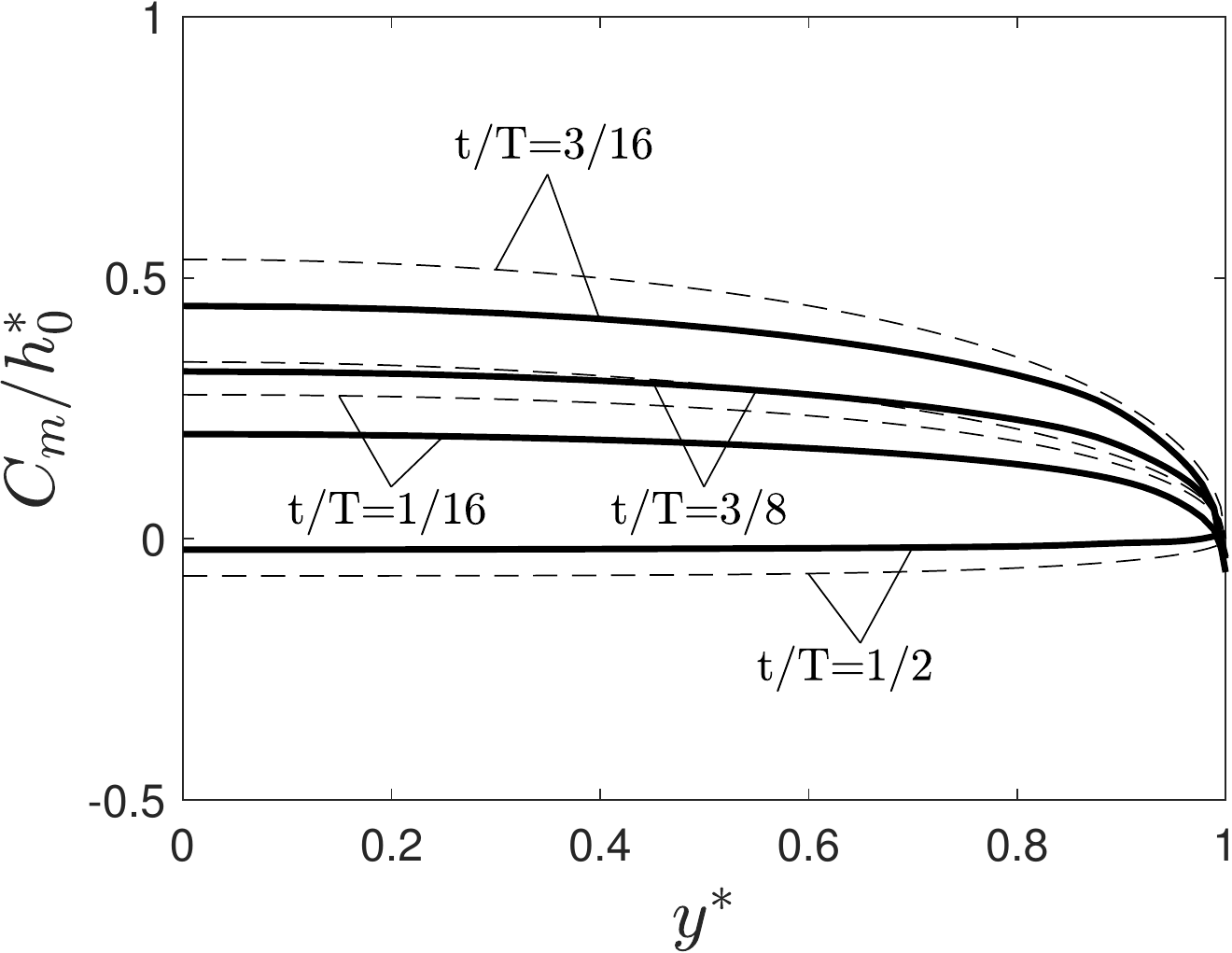}}

	\caption{Lift and moment distributions for rectangular wings oscillating in heave with
	amplitude $h_0^*=0.05$ at $k=0.4$. Moments are taken about the mid-chord.}
	\label{fig:Cl_cm_la}	
\end{figure}

When the assumptions of its derivation are satisfied, ULLT can predict force distributions along
the span of wing. Here, we compare the data obtained from ULLT to the data obtained from CFD.
The lift and moment distributions for the $h_0^*=0.05$ amplitude at 
aspect ratios 1 and 6 are shown in Fig.~\ref{fig:Cl_cm_la}.

Both distributions at both aspect ratios have broadly smooth curves. However, the lift distribution
at the very tip of the wing is non-smooth due separation at the sharp edges of the squared off tips. 
The overestimation of the force amplitude follows that obtained for whole wing forces
found earlier.
At aspect ratio 6, shown in Fig.~\ref{fig:Cl_la_ar6}, ULLT predicts the shape of the
curves very well. The lift
coefficient varies significantly with respect to span only  near the wing
tip, suggesting that the flow is 2D over the majority of the wing.
For the aspect ratio 1 case (Fig.~\ref{fig:Cl_la_ar1}), the prediction is
worse. For low Reynolds number, low amplitude cases,  ULLT
provides a good prediction of force distribution with respect to span.

\begin{figure}
\centering
\begin{tabular}{ >{\centering\arraybackslash}m{0.05\textwidth}  >{\centering\arraybackslash}m{0.45\textwidth}  >{\centering\arraybackslash}m{0.45\textwidth}}
\toprule
$t/T$ & Aspect ratio 6 & Aspect ratio 1\\
\midrule
	$\frac{1}{16}$
	&\includegraphics[height=0.15\textwidth]{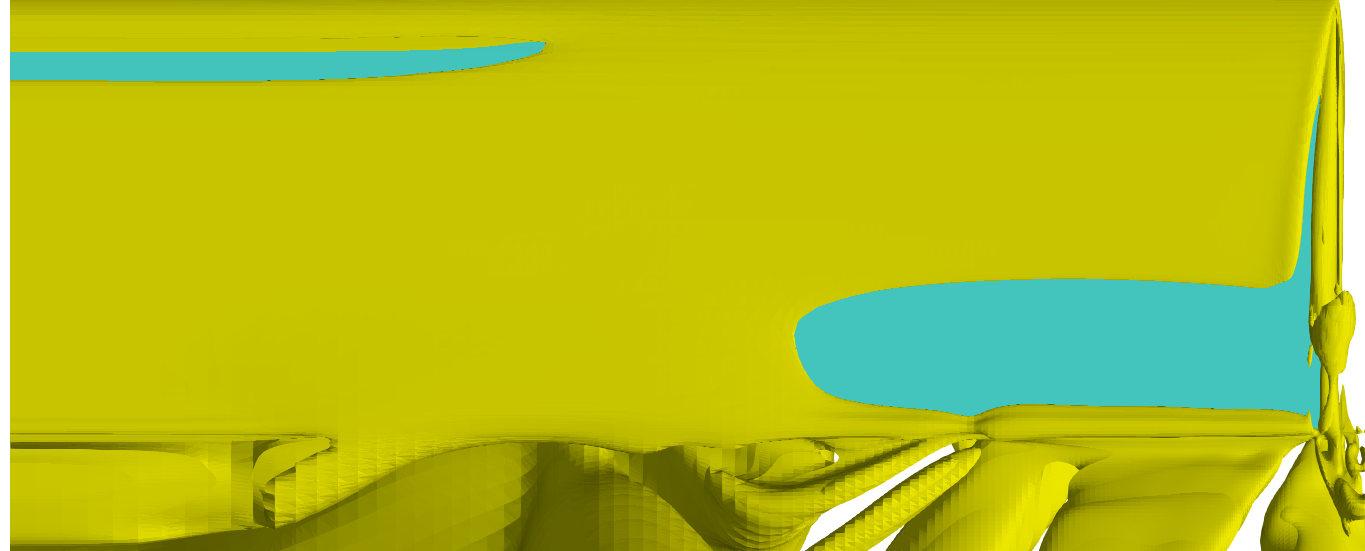}
	&\includegraphics[height=0.15\textwidth]{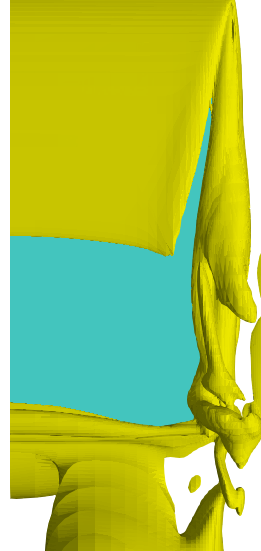}\\
	$\frac{1}{8}$
	&\includegraphics[height=0.15 \textwidth]{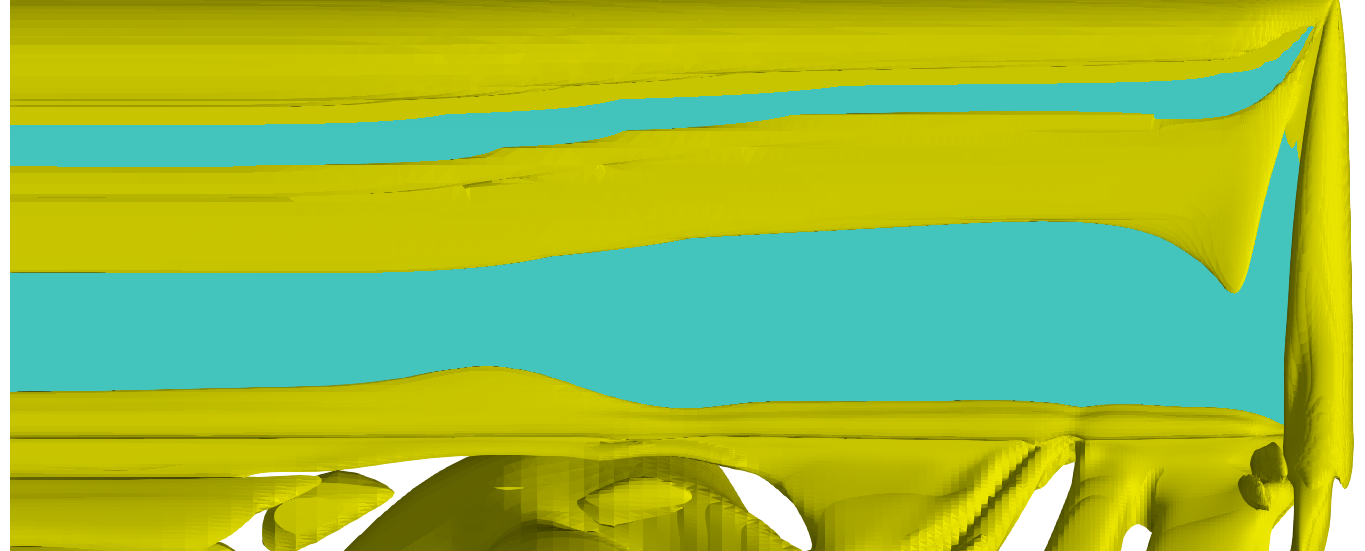}
	&\includegraphics[height=0.15 \textwidth]{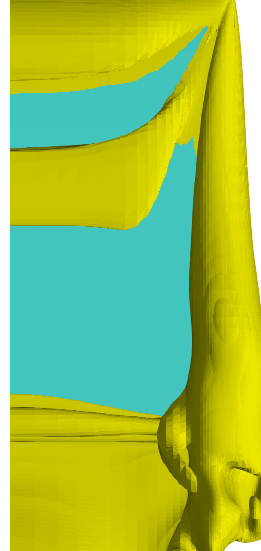}\\
	$\frac{3}{8}$
	&\includegraphics[height=0.15 \textwidth]{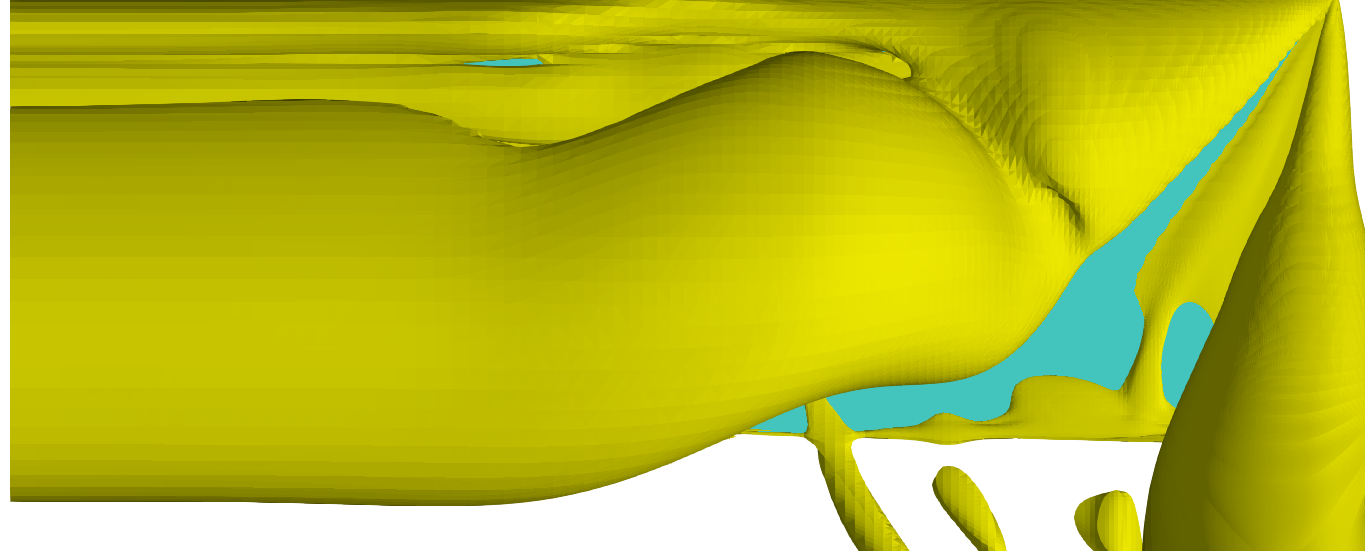}
	&\includegraphics[height=0.15 \textwidth]{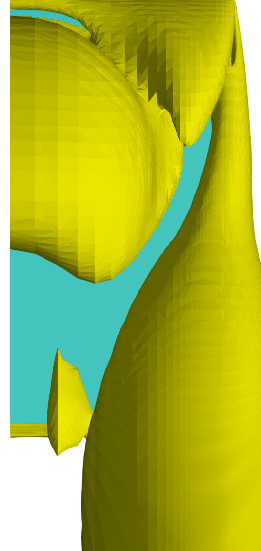}\\
	$\frac{1}{2}$
	&\includegraphics[height=0.15 \textwidth]{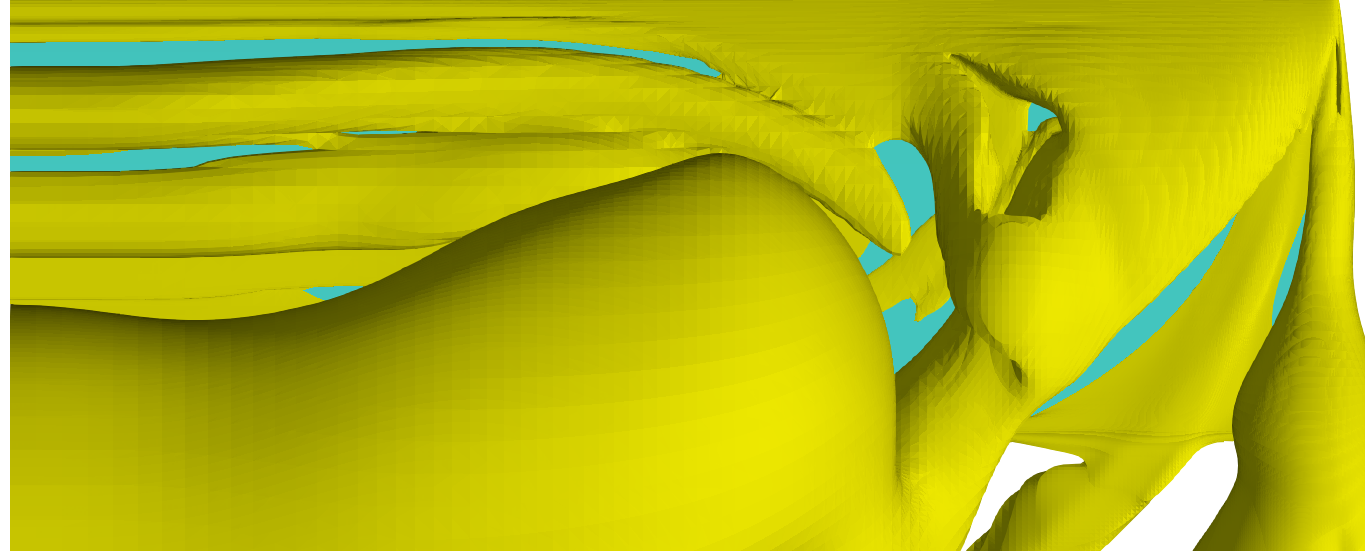}
	&\includegraphics[height=0.15 \textwidth]{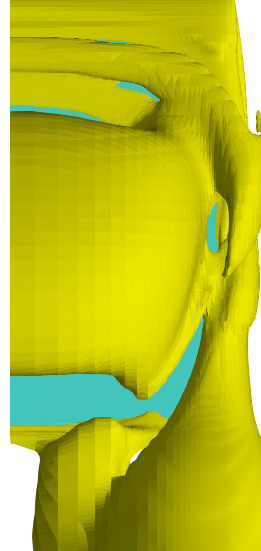}\\
	 
	& \includegraphics[width=0.4 \textwidth]{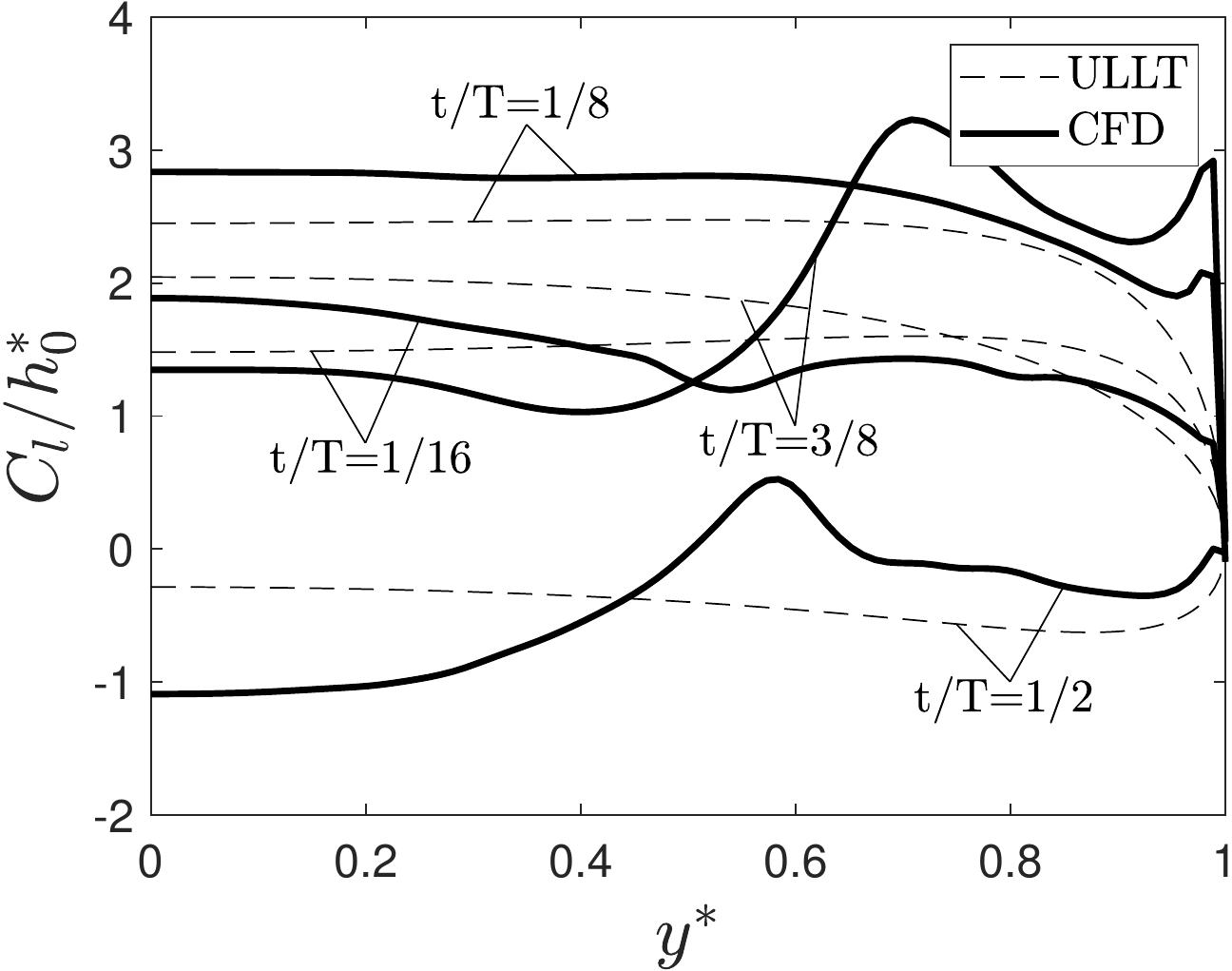}
	& \includegraphics[width=0.4 \textwidth]{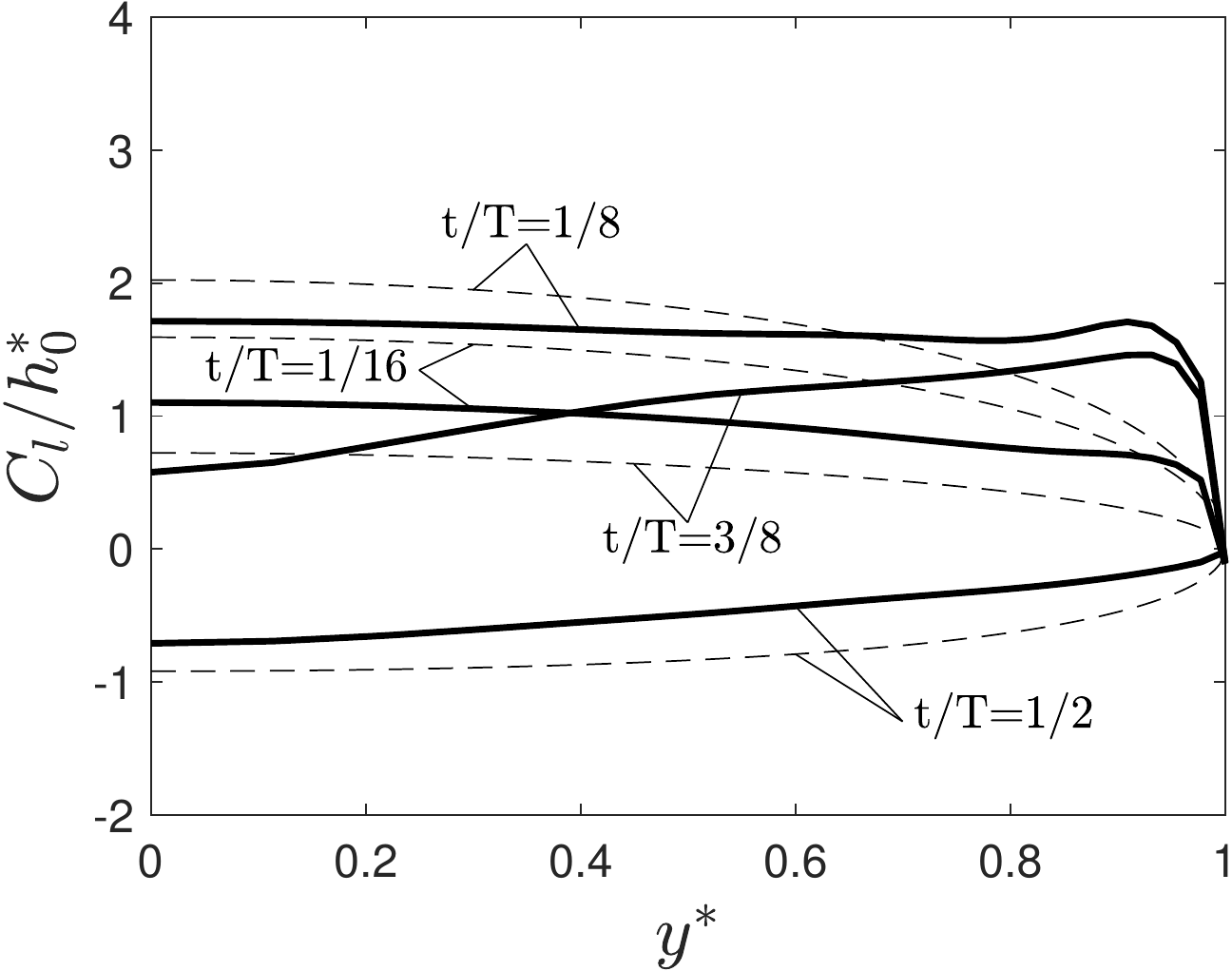} \\
	   
	& \includegraphics[width=0.4\textwidth]{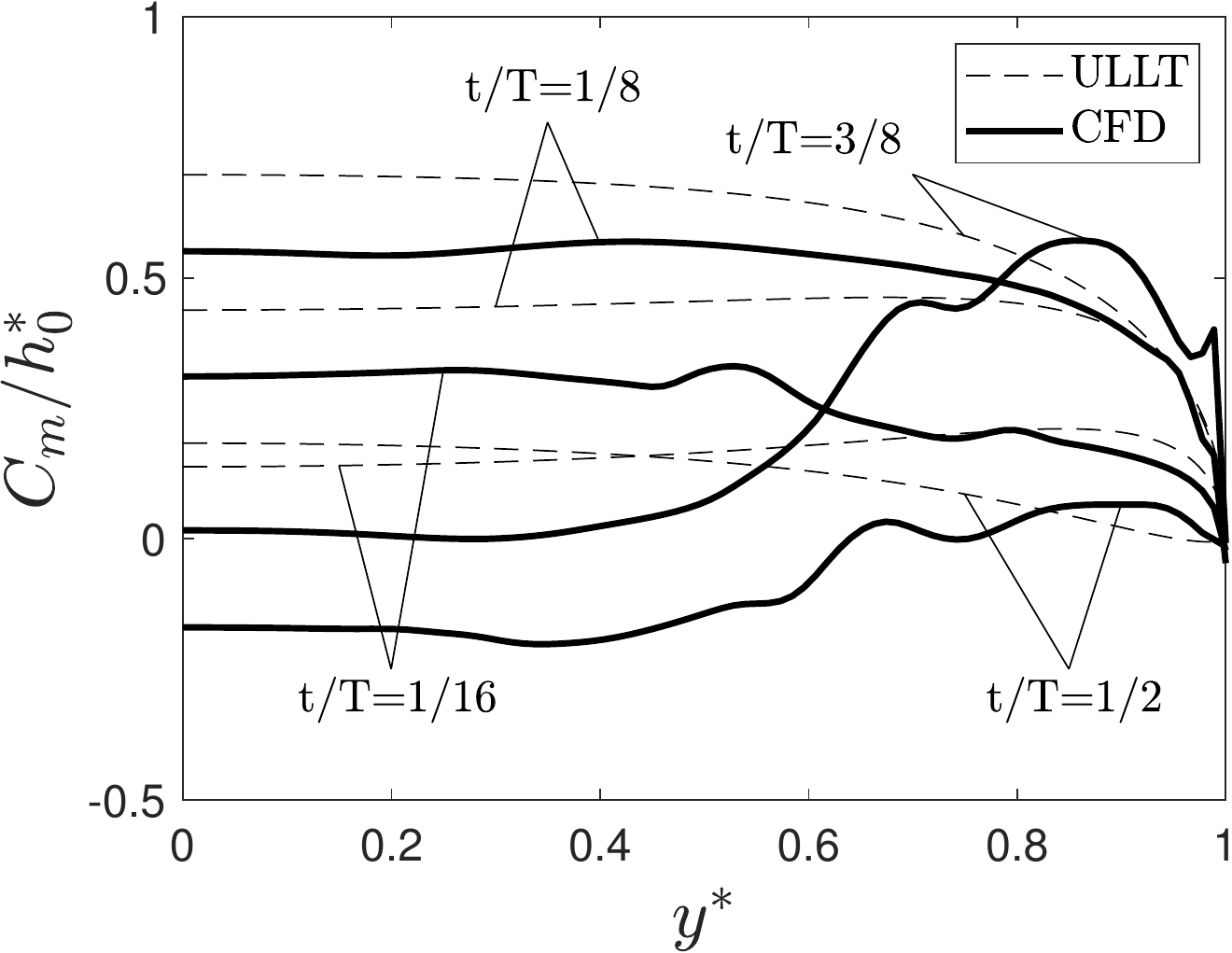}
	& \includegraphics[width=0.4 \textwidth]{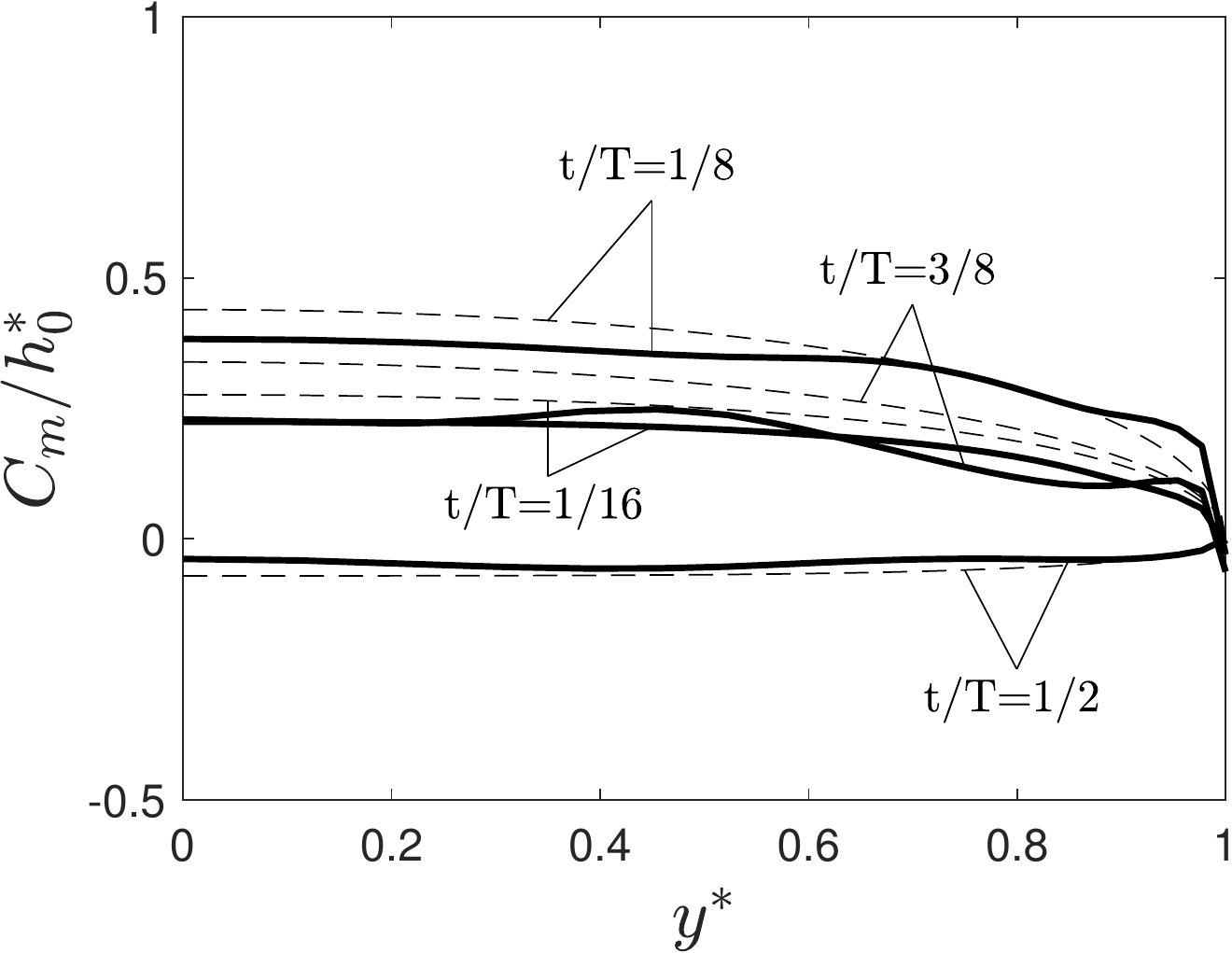} \\
	    	
\bottomrule
\end{tabular}
		\caption{Lift distributions, moment distributions and Q criterion 
		iso-surfaces ($Q=1$) for a wing oscillating in heave with amplitude $h_0^*=1$ 
		at $k=0.4$. Moments are taken about the mid-chord. }
		\label{fig:LEVd_distributions}
\end{figure}

Figure~\ref{fig:LEVd_distributions} shows the force distributions for
the large amplitude $h_0^*=1$ cases, along with a visualization
of the LEV obtained by taking a Q criterion iso-surface for $Q=1$. Both the
force and vorticity distributions are shown at times $t/T = \{1/16,
1/8, 3/8, 1/2\}$. These times correspond to the growth of the $h_0^*=1$
LEV shown in Fig.~\ref{fig:wing_centre_vort_plots}. 

At aspect ratio 1, the vortex structures that forms through
the sinusoidal stroke results in an increase in forces near the wing
tips. At the center of the wing, the $C_l$ and $C_m$ are similar to that
predicted by unsteady lifting-line theory.
ULLT combined with the LESP criterion predicted that LEV shedding would 
occur in these high amplitude cases, as shown in Figure~\ref{fig:a0_wing}.
The LESP criterion does correctly predict LEV shedding in
the broadest sense at low aspect ratio. Notably, whilst Fig.~\ref{fig:A0_amplitude}
suggests that LEV shedding does not occur near the wing tip,
significant vortical structures still exist here. 
The 2D flow that ULLT assumes at each point on the wing span is
evidently incorrect around the prominent vortices at the wing tips.
Consequently,
the lift distribution obtained from ULLT does not match that from CFD.

At aspect ratio 6, the leading-edge vortex structure becomes more complex.
This manifests itself in both the $C_m$ and $C_l$ distribution. 
At $t/T=1/8$, a nascent vortex is forming along almost the entire leading
edge. However, by $t/T=3/8$ an large arch structure has been formed.
The feet of the arch stay connected to wing surface, with vortex
structures extending to the very leading-edge corner of the wing.
This arch evolves with time, with the feet of the arch moving
backwards and inwards. The arch structure has a significant impact
on the force distributions. At $t/T=1/16$,  the $C_l$ and $C_m$ curves
are relatively smooth, with some non-smoothness resulting due to the
shedding of an LEV on the return stroke of the previous oscillation. They
are still smooth at $t/T=1/8$ at the beginning of LEV formation, 
and less effected by the previous LEV. However, the large LEV arch
structure formed by $t/T=3/8$ has a significant impact. The impact on 
force distribution depends on the location of the arch. Outside the arch,
near the wing tips, the influence of the arch is to increase lift. Inside the arch,
lift is decreased, possibly resulting in premature negative lift coefficients compared
to the non-LEV shedding case. The effect is strongest when the LEV is closest to the
wing at $t/T=3/8$ compared to $t/T=1/2$. Near the wing tip, the
force distribution from CFD is also influenced by separation at the wing tip.

Again, ULLT combined with LESP predicts that LEV formation will occur,
but does not yield useful information about how the force
distributions are impacted. The difference between the ULLT prediction and
the CFD result at $t/T=1/16$ emphasizes how, despite the fact that ULLT predicts
that $\mathcal{L}(y) < \mathcal{L}_{crit}$ across the entire wing span, the LEV
shed in previous oscillations can impact the lift distribution on the wing.

\begin{figure}
\centering
    \subfigure[Aspect ratio 6]{
	\label{fig:Cl_LEVa_ar6}
    	\includegraphics[width=0.4\textwidth]{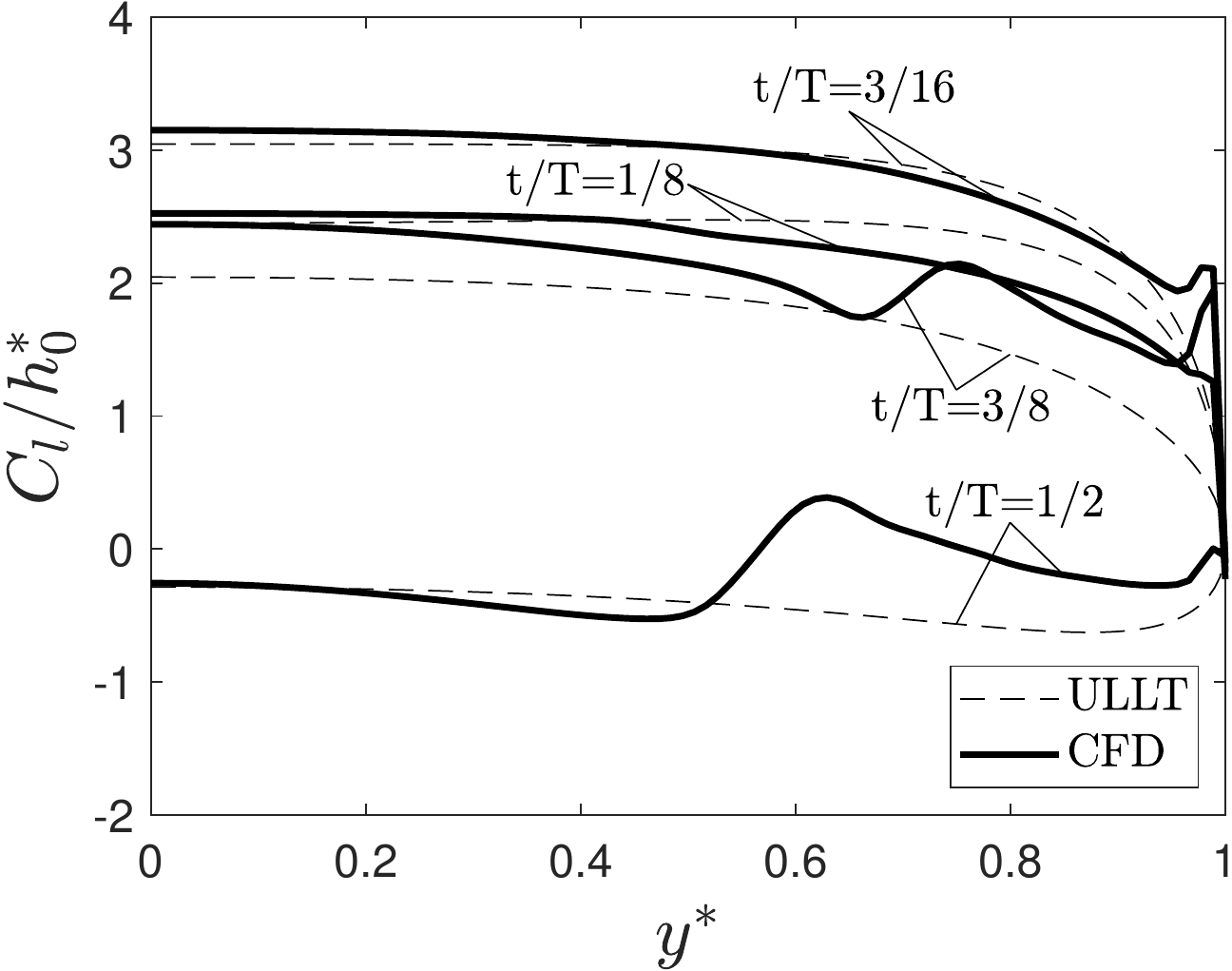}}	
    \subfigure[Aspect ratio 1]{
	\label{fig:Cl_LEVa_ar1}
    	\includegraphics[width=0.4\textwidth]{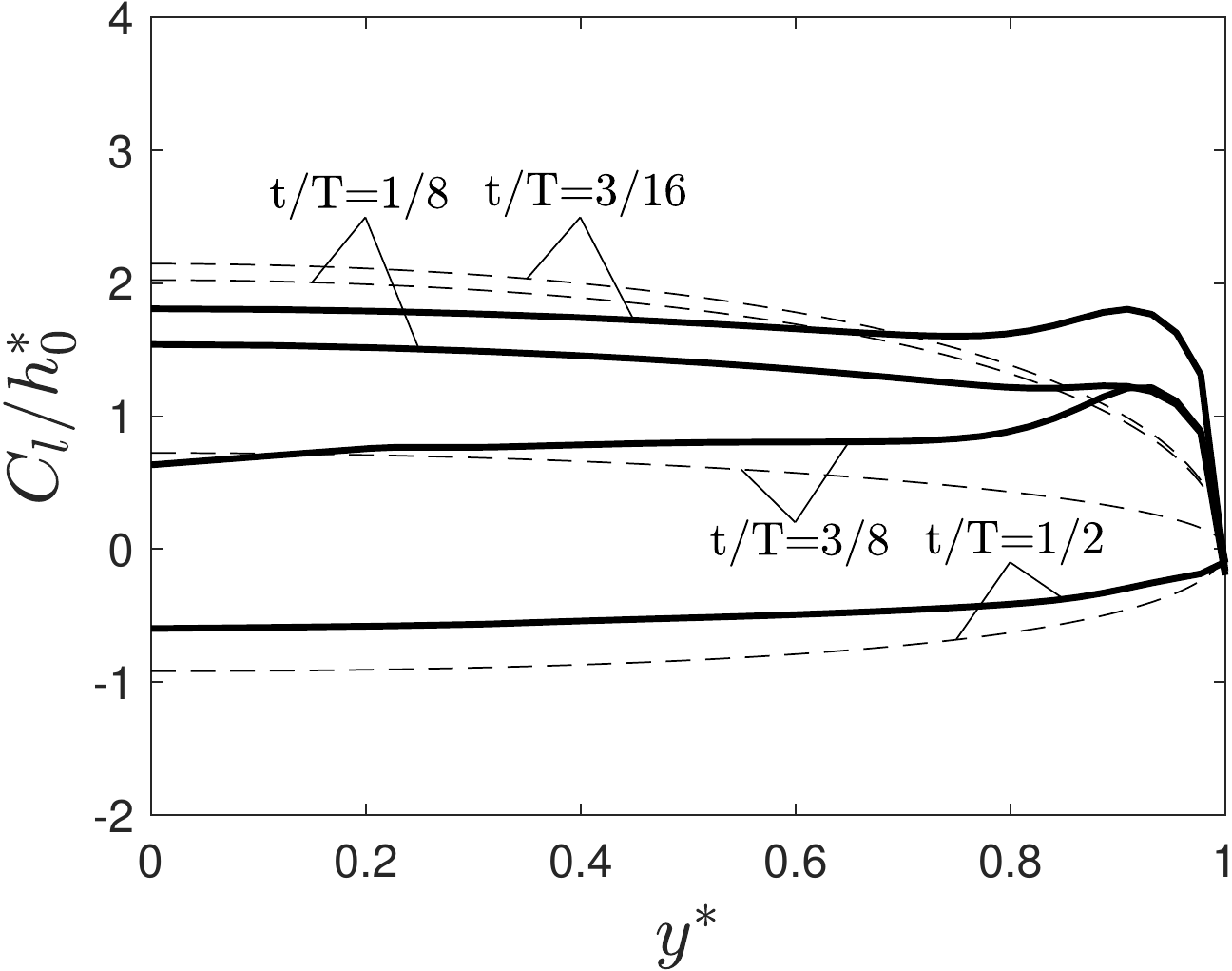}}
    		
	\caption{$C_l$ distribution for rectangular wings oscillating in heave with
	amplitude $h_0^*=0.5$ at $k=0.4$.}
	\label{fig:LEVa_distributions}	
\end{figure}

Finally, the lift distributions for the $h_0^*=0.5$ cases are examined, shown
in Fig.~\ref{fig:LEVa_distributions}. The lift distributions are given at
$t/T=\{1/8,3/16,3/8,1/2\}$, matching the time steps used to study 
the growth of the LEV structure
shown in Fig.~\ref{fig:wing_centre_vort_plots}. 
 
The $C_l$ distribution for the aspect ratio 1 case is shown in Fig.~\ref{fig:Cl_LEVa_ar1}.
The lift distributions obtained from the CFD are similar to that for the
higher amplitude $h_0^*=1$ case (see in Fig.~\ref{fig:LEVd_distributions}). 
The ULLT fails to predict the increased lift near the wing tip found in the CFD
data. The ULLT combined with the LESP criterion predicts LEV shedding in
the center of the wing (see Fig.\ref{fig:A0_amplitude}), with the critical
value of the leading edge suction parameter only just being exceeded.
The effects of the LEV
on the $C_l$ distribution appear to be negligible. Consequently, the
error in the ULLT prediction of forces stems instead from the tip
vortices instead. 

At aspect ratio 6, shown in Fig.~\ref{fig:Cl_LEVa_ar6}, 
the inability of ULLT to capture the detail of the lift distribution
obtained from CFD is again demonstrated. 
It is accurate at $t/T=1/8$, before LEV formation.
However, like the $h_0^*=1$ case (see Fig.~\ref{fig:LEVd_distributions}), 
the LEV arch eventually creates a recognizable disturbance in the 
lift distribution that the ULLT cannot replicate. 
Due to the lower amplitude of oscillation, ULLT combined with the LESP
criterion predicts that the critical value of leading edge suction is exceeded by
a lesser extend in comparison to the larger amplitude case.
As a result, the resulting LEV is weaker, and the impact of the LEV of the 
lift distribution is diminished. 
The LESP criterion cannot reflect the 3D evolution
of the vortex structure, but the extent to which the critical
LESP is exceeded does broadly predict the extent to which
the predicted force distribution is incorrect.

\section{Conclusions}
\label{sec:conclusions}

An Unsteady Lifting-Line Theory (ULLT) for wings oscillating in heave 
was presented, including a means to find both whole wing forces,
wing force distribution and leading edge suction distribution. This
inviscid, small-amplitude theory was then compared to CFD
results for low Reynolds number regime cases, including ones
with large amplitude oscillations leading to leading edge vortex shedding.

ULLT accurately predicted the whole wing lift and moments 
for rectangular wings. For small amplitude oscillations,
the prediction was very good. As the amplitude increased,
aerodynamic non-linearities caused deviation from the assumed
sinusoidal form of the forces with respect to time. However,
these deviations from the sinusoidal form reduced significantly
as aspect ratio decreased - 3D effects stabilized the leading
edge vortex structure. 

For small amplitude oscillation, the ULLT predicted the load distribution
reasonably well. As oscillation amplitude increased, aerodynamic
non-linearities, resulted in a poor prediction of the wing load distribution.
The prediction became worse as the amplitude increased.

The ULLT, combined with the leading edge suction criterion, could
be used to predict leading edge vortex shedding, but not in detail.
Consequently, it acts as a coarse indicator of the usefulness of
of the ULLT predicted wing load distributions. This indicator 
is less useful at very low aspect ratio due to the importance of
vortical structures caused by separation at the wing tips.

Broadly, we find that inviscid ULLT is indeed very useful 
in approximate prediction of oscillating wings in the low
Reynolds number regime.

\section*{Funding Sources}

The authors gratefully acknowledges the support of the UK
Engineering and Physical Sciences Research Council (EPSRC)
through a DTA scholarship, grant EP/R008035 and the
 Cirrus UK National Tier-2 HPC service at 
 EPCC (http://www.cirrus.ac.uk),
and to the Japan Student Services Organization (JASSO) for 
their generous financial support.

\bibliography{JabRefDatabase,kiran_bibtot}   

\end{document}